\documentclass[aps,prd,reprint,twocolumn,nofootinbib,longbibliography]{revtex4-2}

\usepackage{amsmath,amssymb,mathtools,bm}
\usepackage{slashed}
\usepackage{graphicx}
\usepackage[colorlinks=true,linkcolor=blue,citecolor=blue,urlcolor=blue]{hyperref}
\usepackage[final]{microtype}
\usepackage{mathrsfs}

\newcommand{\ie}{\textit{i.e., }}
\newcommand{\Tr}{\mathrm{Tr}}
\newcommand{\ket}[1]{\left|#1\right\rangle}

\newcommand{\mel}[3]{\left\langle #1\middle|#2\middle|#3\right\rangle}
\newcommand{\braket}[2]{\left\langle #1\middle|#2\right\rangle}
\newcommand{\avg}[1]{\left\langle #1\right\rangle}
\newcommand{\Ah}{\hat A}
\newcommand{\ah}{\hat a}
\newcommand{\ad}{\hat a^{\dagger}{}}
\newcommand{\nh}{\hat n}
\newcommand{\rhoh}{\hat\rho}
\newcommand{\Uh}{\hat U}
\newcommand{\Dh}{\hat D}
\newcommand{\Sh}{\hat S}
\newcommand{\Rh}{\hat R}
\newcommand{\Kh}{\hat K}
\newcommand{\Vh}{\hat{\mathcal V}}

\begin{document}

\title{Nonlinear Compton Scattering in a Quantized Pump Field}

\author{Kenan Qu}
\email{kq@princeton.edu}
\affiliation{Department of Astrophysical Sciences, Princeton University, Princeton, New Jersey 08544, USA}


\date{\today}

\begin{abstract}
We develop a fully quantized theory of nonlinear Compton scattering driven by a single-mode quantum field. Exact quantum-Volkov states retain pump depletion, back-action, and final-state correlations through displaced or squeezed-displaced Fock-state ladders. A finite Fock-state pump produces discrete photon-transfer edges and a terminal spectral cutoff. In the bright, weakly depleted regime, the exact theory reduces to a Wigner-function weighted-average of scattering probabilities evaluated at fixed complex field amplitudes, with ordinary and generalized Bessel functions describing the harmonic structure for circular and linear polarization, respectively. For squeezed coherent light, the squeezing angle controls the high-energy emission through photon-number fluctuations. 
\end{abstract}

\maketitle

\section{Introduction}

Nonlinear Compton scattering (NLC) is an elementary process in strong-field quantum electrodynamics. In the standard Furry-picture formulation, a charged particle interacts nonperturbatively with a prescribed intense electromagnetic plane wave and emits a photon into an otherwise quantized radiation field. The external plane wave is treated as a classical background, and the electron dynamics is described by Volkov states \cite{Volkov1935,BrownKibble1964,NikishovRitus1964,Ritus1985,DiPiazza2012, DiPiazza2018}. This theory underlies the interpretation of laser-driven high-energy photon sources and a series of nonlinear QED experiments, including  the SLAC E144~\cite{Bula1996,Bamber1999}, FACET-II E320~\cite{Cavanagh2023}, and the Astra Gemini experiments~\cite{Cole2018, Poder2018}. In this formulation, the quantum emission is interpreted in terms of multiphoton absorption, but the laser state itself is not part of the Hilbert space and therefore cannot be depleted.

A central question, already considered in early quantum-field treatments of strong-field processes, is how the dynamics change when the intense driving mode is treated as a quantum degree of freedom rather than a prescribed classical background. In 1969, Berson~\cite{Berson1969} obtained exact solutions for an electron in a quantized monochromatic radiation mode. Bergou and Varr\'{o} subsequently developed nonrelativistic and relativistic theories of nonlinear scattering in a quantized radiation field. They showed explicitly that the semiclassical Volkov description emerges in the high-intensity, weak-depletion limit, whereas the fully quantized theory retains photon-number transitions and depletion corrections \cite{BergouVarro1981Nonrel,BergouVarro1981Rel}. Guo and Åberg extended this framework to elliptically polarized quantized fields and applied the resulting quantized Volkov states to multiphoton ionization, further clarifying how the photon-number index in the quantum theory becomes the absorbed-photon order in the classical strong-field expression \cite{GuoAberg1988}. Subsequent work connected these quantized-field constructions to quantum optics and high-harmonic generation, including photon-counting signatures, phase-space descriptions, and quantized strong-field Kramers--Heisenberg formulations \cite{Gonoskov2016,Gombkoto2020,Gorlach2020,Varro2021,Varro2022}. The present relevance of this question is motivated by experimentally accessible intense nonclassical states, including squeezed, thermal, and bright squeezed-vacuum light, as well as recent proposals and demonstrations of quantum-light-driven strong-field phenomena \cite{Gorlach2023,EvenTzur2024PRR,Rasputnyi2024,EvenTzurCohen2024, Qu_PRE_entangle24,Qu_squeezing2025}.

For nonlinear Compton (NLC) scattering, Khalaf and Kaminer recently developed a phase-space formulation for an intense quantum-light drive \cite{KhalafKaminer2023}. Their method represents the incident light by a generalized Glauber--Sudarshan distribution, $P(\alpha,\beta)$, and decomposes the dynamics into branches driven by classical coherent amplitudes. For circular polarization, the inclusive spectrum is expressed in terms of a scaled Husimi $Q$ function evaluated at the effective field amplitudes selected by the nonlinear Compton channels. This representation is compact and transparent for inclusive spectra, but the coherent-state branch decomposition and subsequent reduction to the $Q$ function describe the pump through a quasistatistical phase-space distribution. Consequently, phase-sensitive coherences, conditional final pump states, pump depletion, and pump--electron--emission entanglement are not retained as explicit dynamical degrees of freedom. The formulation predicts broader spectra for thermal and bright squeezed-vacuum driving fields than for a coherent drive of the same average intensity \cite{KhalafKaminer2023}.

Here, we develop a full quantum formulation that retains the pump as a quantum subsystem throughout the Furry-picture construction. The electron state is dressed by the pump field in the Fock-state basis, and the NLC amplitude explicitly connects an incoming dressed-ladder rung $n$ to an outgoing rung $m$. This construction retains the finite-occupation transition matrix elements, pump depletion, and correlations between the emitted particles and the residual pump, while also permitting conditional measurements or postselection of the final pump state. These quantities are absent from the classical Volkov theory, in which the pump is a prescribed classical background. Although pump depletion and back-action have been studied by using different initial and final coherent states~\cite{SeiptHeinzlMarklundBulanov2017, IldertonSeipt2018}, these descriptions retain a classical treatment of the coherent driving field. 
These quantum fields are also not resolved by the generalized-$P$ formulation of Ref.~\cite{KhalafKaminer2023}, in which the emission spectrum is obtained from a constant coherent-field basis. The Fock-space formulation is therefore most useful for finite pump occupations, appreciable photon transfer, and pump-resolved observables. As an example, we show that a finite Fock-state pump leads to discrete photon-transfer edges and a terminal spectral cutoff imposed by the bounded pump occupation.

For bright pumps with weak depletion, direct summation over the dressed Fock ladder becomes computationally inefficient. We therefore derive a Wigner--Weyl reduction of the emission probability. It is particularly convenient for Gaussian states, including squeezed coherent light, whose Wigner functions are regular distributions over the physical complex amplitude~\cite{Agarwal_2012}. In the high-occupation, undepleted regime, the spectrum reduces to a Wigner-weighted average of coherent NLC probabilities, thereby retaining the initial amplitude and phase fluctuations while neglecting pump depletion. The circularly and linearly polarized amplitudes are expressed in terms of ordinary and generalized Bessel functions, respectively. When applied to a bright squeezed coherent pump, the Wigner formulation shows how the relative squeezing angle and magnitude reshape the photon-number and intensity fluctuations, thereby modifying the harmonic spectrum and high-energy yield. The emission spectrum reduces to the classical result in the absence of squeezing. Compared with the generalized-$P$-function framework of Ref.~\cite{KhalafKaminer2023}, the Wigner formulation produces the same spectra while providing a direct interpretation and numerical implementation in terms of physical quadrature distributions. 

Note that a complementary form of quantum-optical control is obtained by engineering the modes into which radiation is emitted. In Ref.~\cite{DiPiazzaQu2026PRL}, the electron is driven by a prescribed intense plane wave, while selected emission modes are prepared in a squeezed vacuum. The resulting modification of vacuum correlations enhances or suppresses the NLC probability depending on the squeezing angle. In the present work, by contrast, the engineered quantum state is that of the driving pump, which is retained as a dynamical quantum subsystem.

The presentation is self-contained and retains the main steps connecting the quantized pump mode, the quantum-Volkov state, and the resulting scattering amplitudes. Section~\ref{sec:hamiltonian} defines the
Hamiltonian and the Furry-picture split, and Sec.~\ref{sec:qvolkov} derives
the operator-valued Volkov states for circular and linear monochromatic
pumps.  Section~\ref{sec:amplitude} constructs the exact state-resolved
scattering amplitude.  The discrete formulation
is illustrated for a Fock-state pump in Sec.~\ref{sec:fock-spectrum}.
Section~\ref{sec:phase-space-evaluation} then derives the undepleted
Wigner--Weyl reduction and identifies its range of validity.
Section~\ref{sec:scs_pump} applies that reduction to squeezed coherent
driving and relates the squeezing-angle dependence of the photon statistics
to the harmonic spectrum and high-energy yield. Section~\ref{sec:conclusion} presents the conclusions and discussion.

\section{Hamiltonian and the Furry picture}
\label{sec:hamiltonian}

We first outline the standard Furry-picture framework, following Refs.~\cite{Volkov1935,BrownKibble1964, NikishovRitus1964,Ritus1985, DiPiazza2012,DiPiazza2018}.
We consider a relativistic electron interacting with a single-mode electromagnetic field. 
In natural units, $\hbar=c=1$, and with the metric
$g^{\mu\nu}=\mathrm{diag}(1,-1,-1,-1)$, the electron--field system is described by the Hamiltonian
\begin{equation}
    H = H_{\rm EM} + H_{\rm Dirac} + H_{\rm int}.
\end{equation}
The three terms describe the electromagnetic field, the Dirac electron, and their interaction, respectively.

The electromagnetic field, $\Ah^\mu = \Ah_L^\mu+\Ah_{\rm rad}^\mu$, is separated into the pump and radiation fields because they are treated differently. The highly occupied pump mode, $\Ah_L^\mu(\phi)$, can strongly modify the electron motion and is therefore treated nonperturbatively. The radiation field, $\Ah_{\rm rad}^\mu(x)$, contains the weakly occupied emitted modes and is treated perturbatively. The latter has the standard mode expansion
\begin{multline}
    \Ah_{\rm rad}^\mu(x)=\sum_{\varsigma}\int\frac{d^3k'}{(2\pi)^3\sqrt{2\omega'}} \\
    \times \left[\epsilon_{k'\varsigma}^\mu\hat b_{\mathbf k'\varsigma}e^{-ik'\cdot x}
+\epsilon_{k'\varsigma}^{\mu *}\hat b_{\mathbf k'\varsigma}^\dagger e^{ik'\cdot x}\right],
\label{eq:Ah}
\end{multline}
where $k'^\mu=(\omega',\mathbf k')$ is the four-momentum of a radiation mode and $\omega'=|\mathbf k'|$. The index $\varsigma$ labels its polarization, and $\epsilon_{k'\varsigma}^\mu$ is the corresponding polarization four-vector. We use $\ah$ for the pump-mode operator and $\hat b_{\mathbf k'\varsigma}$ for the emitted-radiation operators. The factor $1/\sqrt{2\omega'}$ gives the standard relativistic normalization. The unprimed four-momentum $k^\mu=(\omega,\mathbf k)$ denotes the strong quantized pump.
In state-resolved scattering amplitudes, $\sigma$ and $\sigma'$ label the initial and final electron spins, $\varsigma$ labels the emitted-photon polarization, and $n$ and $m$ label the incoming and outgoing dressed-ladder rungs.

The Hamiltonian $H_{\rm EM}$ describes both the pump mode and the radiation continuum:
\begin{equation}
 H_{\rm EM}=\omega\left(\ad\ah+\frac12\right) + \sum_{\varsigma}\int\frac{d^3k'}{(2\pi)^3}\,\omega'\,
 \hat b^\dagger_{\mathbf k'\varsigma}\hat b_{\mathbf k'\varsigma}.
\label{eq:HL}
\end{equation}
Here, $\ah$ annihilates a photon in the single-mode pump, and $\hat b_{\mathbf k'\varsigma}$ annihilates a radiation photon with wave vector $\mathbf k'$ and polarization $\varsigma$. The prime distinguishes a radiation mode from the pump mode.

In QED, a relativistic electron state $\psi(x)$ in an electromagnetic four-potential obeys the Dirac equation. In the Hamiltonian formulation, the free Dirac Hamiltonian is
\begin{equation}
 H_{\rm Dirac}=\int d^3x\,\hat\psi^\dagger(x)
 \left[-i\bm\alpha\cdot\bm\nabla+\beta m\right]\hat\psi(x),
 \label{eq:HDirac-x}
\end{equation}
where $\bm\alpha=\gamma^0\bm\gamma$ and $\beta=\gamma^0$. 
The Dirac field is expanded in electron and positron creation and annihilation operators as
\begin{multline}
\hat\psi(x)
=
\sum_{\sigma}
\int
\frac{d^3p}{(2\pi)^3}
\frac{1}{\sqrt{2E_{\mathbf p}}}
\\
\times
\left[
\hat c_{\mathbf p\sigma}
u_\sigma(p)
e^{-ip\cdot x}
+
\hat d_{\mathbf p\sigma}^{\dagger}
v_\sigma(p)
e^{ip\cdot x}
\right],
\label{eq:dirac_field_expansion}
\end{multline}
where $\hat c_{\mathbf p\sigma}$ annihilates an electron with momentum $\mathbf p$ and spin $\sigma$, while $\hat c_{\mathbf p\sigma}^\dagger$ creates such an electron.  Similarly, $\hat d_{\mathbf p\sigma}$ annihilates a positron and $\hat d_{\mathbf p\sigma}^\dagger$ creates a positron.  The adjoint field is $\widehat{\bar\psi}(x)=\hat\psi^\dagger(x)\gamma^0$, and the nonzero anticommutation relations are
\begin{equation}
\left\{
\hat c_{\mathbf p\sigma},
\hat c_{\mathbf p'\sigma'}^\dagger
\right\}
=
\left\{
\hat d_{\mathbf p\sigma},
\hat d_{\mathbf p'\sigma'}^\dagger
\right\}
=
(2\pi)^3
\delta^{(3)}(\mathbf p-\mathbf p')
\delta_{\sigma\sigma'} .
\label{eq:fermion_anticommutators}
\end{equation}
Thus, $ H_{\rm Dirac}$ can be rewritten as a sum over  electron and positron energies
\begin{equation}
 H_{\rm Dirac}=\sum_{\sigma}\int\frac{d^3p}{(2\pi)^3}\,E_{\mathbf p}
 \left(\hat c_{\mathbf p\sigma}^\dagger\hat c_{\mathbf p\sigma}
 +\hat d_{\mathbf p\sigma}^\dagger\hat d_{\mathbf p\sigma}\right),
 \label{eq:HDirac-p}
\end{equation}
where $E_{\mathbf p}=\sqrt{\mathbf p^2+m^2}$.

The electron--field interaction couples the Dirac current to the electromagnetic field. Classically, the field accelerates the electron through the Lorentz force. Quantum mechanically, the coupling transfers energy and momentum between the electron and photon modes. Exchange with the strong pump can involve many photons and must be included to all orders, whereas emission into the radiation continuum is treated perturbatively. Separating these fields gives
\begin{equation}
 H_{\rm int}=e\int d^3x\,\widehat{\bar\psi}(x)\gamma_\mu
 \left[\Ah_L^\mu(\phi)+\Ah_{\rm rad}^\mu(x)\right]\hat\psi(x).
\label{eq:Hint}
\end{equation}
Here $\Ah_L^\mu(\phi)$ and $\Ah_{\rm rad}^\mu(x)$ are written with different arguments because
they play different roles.  The strong driving mode is taken to be a single
monochromatic plane-wave mode, so its spacetime dependence is only through
the light-front phase $\phi = k\cdot x $. 
On the other hand, $\Ah_{\rm rad}^\mu(x)$ denotes the full continuum of perturbative
radiation modes and is therefore kept as a general field operator of spacetime. 
In the following, we use the slash notation for contraction with Dirac matrices \ie
$\slashed{\Ah}\equiv\gamma_\mu\Ah^\mu$ and
$\slashed{k}\equiv\gamma_\mu k^\mu$.

Because the pump strongly modifies the electron dynamics, the relevant eigenstates are not those of $H_{\rm Dirac}$ alone but electron--pump dressed states that include the pump photon-number structure. We thus adopt the Furry picture, in which the dressed-state basis is determined by
\begin{equation}
 H_F=H_{\rm Dirac}+ \omega\left(\ad\ah+\frac12\right)
 +e\int d^3x\,\widehat{\bar\psi}\slashed{\Ah}_L\hat\psi.
 \label{eq:HF}
\end{equation}
In this basis, emission into modes outside the pump is described as a transition between dressed states accompanied by the creation of a radiation photon through
\begin{equation}
 H_{\rm rad}=e\int d^3x\,\widehat{\bar\psi}\slashed{\Ah}_{\rm rad}\hat\psi,
 \label{eq:Hprime}
\end{equation}
where both operators $\Ah_{\rm rad}$ and $\hat\psi$ can be written as annihilation and creation operators [see Eq.~\eqref{eq:Ah} and \eqref{eq:dirac_field_expansion}]. 
The total Hamiltonian in the Furry picture is thus
\begin{equation}
 H =  H_F + \sum_{\varsigma}\int\frac{d^3k'}{(2\pi)^3}\,\omega'\,
 \hat b^\dagger_{\mathbf k'\varsigma}\hat b_{\mathbf k'\varsigma} + H_{\rm rad}. 
 \label{eq:Htotal}
\end{equation}
The three terms describe the pump-dressed electron, the radiation field, and the interaction responsible for emission.

The Hamiltonian $H_F$ plays the role of the unperturbed Hamiltonian for the scattering problem.  It already contains the dominant, strong electron--pump interaction, while $H_{\rm rad}$ describes the comparatively weak coupling to the emitted radiation modes.  In the classical limit, where the pump has a large photon number and negligible relative depletion, the pump operator can be replaced by its coherent expectation value.  The dressed states then reduce to the usual Volkov-type states of the classical Furry picture.  In the quantum-pump formulation, however, the pump remains an operator, so the scattering process can change the pump photon number and can entangle the electron, the pump mode, and the emitted radiation.

The Furry picture is analogous to the dressed-state formulation of the Jaynes--Cummings model. When an atom is strongly coupled to a quantized radiation mode, the interaction is incorporated into the unperturbed Hamiltonian, whose eigenstates are superpositions of bare atomic and photonic states. Transitions are consequently evaluated between these dressed eigenstates rather than between bare product states.
The Furry Hamiltonian plays an analogous role in strong-field QED, but the
electron is dressed by a relativistic quantized radiation mode, producing a Volkov state,  similar to the atom--cavity dressed states. However, the dressing field is the strong pump rather than the emission field as in the atom--cavity dressed states.

NLC is induced perturbatively by $\Ah_{\rm rad}^\mu$, which connects the initial and final Volkov states. 
Although the fundamental QED vertex $e\widehat{\bar\psi}\gamma_\mu\Ah^\mu\hat\psi$ remains linear in the radiation field, the pump interaction is included to all orders in the dressed electron states. 
In this sense, the Furry picture converts the linear QED vertex into an
effective nonlinear scattering problem between dressed states.  The
nonlinear dependence is encoded in the dressed-state energies involving multiple pump-photon pathways. This is the quantum counterpart of the classical Volkov nonlinearity in strong-field QED, where the electron wave function depends nonperturbatively on the laser amplitude and nonlinear Compton scattering contains harmonics of the driving field.

\section{Quantum Volkov state}
\label{sec:qvolkov}

The classical Volkov solution satisfies the Dirac equation in a prescribed classical plane-wave field. The laser is a fixed $c$-number background rather than a quantum mode with finite occupancy. Replacing the classical field $\slashed{A}$ with the operator $\slashed{\Ah}$ gives the operator-valued Dirac equation $[i \slashed\partial-e\slashed{\Ah}(\phi)-m]\Psi_{p\sigma}(x)=0$, where $\Psi_{p\sigma}(x)$ is an electron spinor whose coefficients act on the pump Hilbert space.

The Volkov state driven by a quantum-state pump takes the form
\begin{equation}
 \Psi_{p\sigma}(x)=\frac{e^{-ip\cdot x}}{\sqrt{2p^0}}
 \left[1+\frac{e\slashed k\slashed{\Ah}_L(\phi)}{2k\cdot p}\right]
 \Uh_p(\phi)u_\sigma(p),
 \label{eq:qVolkov}
\end{equation}
where $u_\sigma(p)$ is the free Dirac spinor for an electron with momentum $p$
and spin $\sigma$. The evolution operator $\Uh_p$ can entangle the pump and electron and satisfies
\begin{equation}
  i \frac{ d }{ d \phi}\Uh_p(\phi)=\hat h_p(\phi)\Uh_p(\phi),
 \label{eq:Up-diff}
\end{equation}
with
\begin{equation}
 \hat h_p(\phi)=\frac{e\,p\cdot\Ah_L(\phi)}{k\cdot p}
 -\frac{e^2:\Ah_L^2(\phi):}{2k\cdot p} .
 \label{eq:hp}
\end{equation}
The normal ordering removes the zero-photon contribution and fixes the vacuum as the undressed reference. Without normal ordering, the additional $c$-number term can instead be absorbed into the reference quasienergy (equivalently, the vacuum dressing convention).

The formal solution for the evolution operator is
\begin{equation}
 \Uh_p(\phi)=\mathcal T\exp\left[- i \int_0^\phi d \varphi\,\hat h_p(\varphi)\right], 
 \label{eq:Texp}
\end{equation}
where $\mathcal T$ denotes time ordering.
This equation is the main structural difference from the classical Volkov state.  For a classical laser, the exponent in Eq.~\eqref{eq:hp} is a number, and all factors commute.  For a quantum pump, $\ah$ and $\ad$ do not commute, so the time ordering cannot be ignored.



To evaluate the ordered exponential, we note that the phase dependence of $\hat h_p(\phi)$ is generated by the number operator $\nh=\ad\ah$ for a monochromatic mode, \ie $e^{i\phi\nh}\ah^r e^{-i\phi\nh}=e^{-ir\phi}\ah^r$, and $e^{i\phi\nh}\ad^{,r}e^{-i\phi\nh}=e^{ir\phi}\ad^{,r}$. The unitary transformation $\hat h_p(\phi)=e^{i\phi\nh}\bar h_p e^{-i\phi\nh}-\nh$ then removes the ordering operation and gives
\begin{equation}
\hat U_p(\phi)
=
e^{i\phi\nh}
e^{-i\phi\bar h_p}.
\label{eq:rotating_hbar_solution}
\end{equation}
Indeed,
\begin{align}
i\frac{d}{d\phi}
\left(
e^{i\phi\nh}e^{-i\phi\bar h_p}
\right)
&=
\left[
-\nh
+
e^{i\phi\nh}
\bar h_p
e^{-i\phi\nh}
\right]
\hat U_p(\phi)
\nonumber\\
&=
\hat h_p(\phi)\hat U_p(\phi).
\end{align}

\subsection{Circularly polarized pump}

To obtain the explicit quantum-Volkov state, we evaluate $\bar h_p$ for the specified pump polarization.
For circular polarization (CP), $\Ah_L^\mu(\phi)=A_0\left(\epsilon^\mu\ah e^{-i\phi}
 +\epsilon^{\mu *}\ad e^{i\phi}\right)$, with $\epsilon^2=0$ and $\epsilon\cdot\epsilon^*=-1$. Hence the normal-ordered field invariant, ${:\Ah_L^2(\phi):}=-2A_0^2\nh$, contains only the number operator.  The phase-independent
generator has the form
\begin{equation}
\bar h_p=\Delta_p\nh+g_p\ah+g_p^*\ad,
\end{equation}
where $\Delta_p=1+\Omega_p$ with the classical intensity parameter $\Omega_p = \frac{e^2A_0^2}{k\cdot p}$, and the linear coupling $g_p=\frac{eA_0\,p\cdot\epsilon}{k\cdot p}$. 
It is diagonalized by the displacement operator $\hat D(\eta_p)=\exp(\eta_p\ad-\eta_p^*\ah)$:
\begin{equation}
\bar h_p
=
\hat D^\dagger(\eta_p)
\Delta_p\nh
\hat D(\eta_p)
-
\frac{|g_p|^2}{\Delta_p}, 
\end{equation}
where  $\eta_p=g_p^*/\Delta_p$.

Because this generator governs the phase evolution of the dressed electron, its exact eigenstates define the intrinsic dressed-ladder basis independently of the prepared pump state. The eigenstates of $\bar h_p$ are displaced Fock states~\cite{Agarwal_2012},
\begin{equation}
\ket{n;p}_{\rm CP} = \Dh^\dagger(\eta_p)\ket{n},
\label{eq:CP_eigenstates}
\end{equation}
where $n = 0, 1, 2, \dots$. The corresponding eigenvalues are
\begin{equation}
\varepsilon_n^{(p)} = \Delta_p n - \frac{|g_p|^2}{\Delta_p}.
\label{eq:CP_eigenvalues}
\end{equation}
These eigenstates and eigenvalues enforce discrete multiphoton energy--momentum conservation in the transition amplitudes. The quantum-Volkov evolution operator is
\begin{align}
 \Uh_p(\phi) &= e^{i\phi\nh}
  e^{i|g_p|^2\phi/\Delta_p}
 \Dh^\dagger(\eta_p)e^{-i\Delta_p\phi\nh}\Dh(\eta_p) \nonumber \\
 &= e^{i\chi_p(\phi)}
 \Dh\!\left[\alpha_p(\phi)\right]
 \Rh\!\left[\Omega_p\phi\right]
 \label{eq:Uc-final}
\end{align}
with
\begin{align}
 \Rh(\theta)&=e^{-i\theta\nh},\\
 \alpha_p(\phi)&=\eta_p e^{i\phi}\left(e^{-i\Delta_p\phi}-1\right),\label{eq:alpha-circ}\\
 \chi_p(\phi)&=\frac{|g_p|^2}{\Delta_p}\phi-\frac{|g_p|^2}{\Delta_p^2}\sin(\Delta_p\phi).
\end{align}

For a head-on collision geometry, the incident electron momentum has no transverse component along the pump polarization, \ie $g_p\propto p\cdot \epsilon=0$. The quantum-Volkov dressing of a CP pump then reduces to a simple phase rotation
\begin{equation}
    \Uh_p(\phi) = \Rh\!\left[\Omega_p\phi\right]
     =  e^{-i\Omega_p\phi\nh}. \label{eq:U_CPheadon}
\end{equation}


Figure~\ref{fig:qvolkov-phase-space} shows the phase-space representation of the dressed pump state $\ket{\Psi_p(\phi)}_L=\Uh_p(\phi)\ket{\Psi}_L$. Its Wigner quasiprobability distribution $W(X, P)$ represents the exact quantum state on a real two-dimensional plane using symmetric operator ordering~\cite{Agarwal_2012}. Figure~\ref{fig:qvolkov-phase-space}(a) shows the Wigner function for a circularly polarized pump. The axes are the dimensionless pump quadratures associated with the Hermitian operators $\hat X=\frac{\hat a+\hat a^\dagger}{\sqrt{2}}$ and $\hat P=\frac{\hat a-\hat a^\dagger}{i\sqrt{2}}$. The photon number is related to the radial coordinate by the exact operator identity $\hat n=\hat a^\dagger \hat a =\frac{1}{2}\left(\hat X^2+\hat P^2-1\right)$. Displacement of the Wigner distribution from the origin represents the coherent photon population, while its shape describes the quantum quadrature fluctuations. For CP, the quantum-Volkov dressing primarily displaces and rotates the Wigner distribution.

\subsection{Linearly polarized pump}

For linear polarization (LP), we take a real transverse polarization vector normalized by $\epsilon^2=-1$. The normal-ordered term is
$:\Ah_L^2(\phi):=A_0^2\epsilon^2
 \left(\ah^2e^{-2i\phi}+\ad^2e^{2i\phi}+2\nh\right)$. Thus the generator $\bar h_p$ contains quadratic terms
\begin{equation}
\bar h_p
=
\Delta_p\nh
+
g_p\ah
+
g_p^*\ad
+
\frac{1}{2}
\left(
\Omega_p\hat a^{\dagger2}
+
\Omega_p\hat a^2
\right).
\label{eq:hbar_linear_explained}
\end{equation}
Its diagonalization requires two Gaussian transformations. For a real LP polarization, $g_p$ is real. First, the displacement operator $\hat D(\eta_p)$ removes the linear terms for $\eta_p=-g_p/(1+2\Omega_p)$. A subsequent Bogoliubov transformation~\cite{Agarwal_2012}, $\hat S(\zeta_p)=\exp\{[\zeta_p^*\hat a^2-\zeta_p\hat a^{\dagger 2}]/2\}$, diagonalizes the quadratic part, where $\zeta_p=r_p e^{i\theta_p}$. Because $\Omega_p$ is real and positive, $\theta_p=0$, and $\tanh(2r_p)=\Omega_p/(1+\Omega_p)$.

Applying these transformations, the generator takes a diagonal form
\begin{equation}
\bar h_p=\Dh(\eta_p)\Sh(\zeta_p)\big(\varpi_p\nh+C_p\big)\Sh^\dagger(\zeta_p)\Dh^\dagger(\eta_p),
\end{equation}
where
\begin{equation}
\begin{aligned}
\varpi_p&=\sqrt{1+2\Omega_p}=e^{2r_p},\\
C_p&=-\frac{g_p^2}{1+2\Omega_p}-\varpi_p\sinh^2r_p.
\end{aligned}
\label{eq:LP-spacing-shift}
\end{equation}
Here $\varpi_p$ determines the dressed energy spacing and $C_p$ is the scalar shift generated by the displacement and Bogoliubov transformations.
The eigenstates of this quadratic generator are squeezed-displaced number states
\begin{equation}
\ket{n;p}_{\rm LP} = \Dh(\eta_p)\Sh(\zeta_p)\ket{n},
\label{eq:LP_eigenstates}
\end{equation}
with the corresponding eigenvalues
\begin{equation}
\varepsilon_n^{(p)} = \varpi_p n + C_p,
\label{eq:LP_eigenvalues}
\end{equation}
where $n = 0, 1, 2, \dots$. Because the squeezed quadrature is aligned with the local coherent displacement, the linearly polarized quantum-Volkov dressing produces amplitude squeezing of the dressed pump component.

From the diagonalized generator, the Volkov phase evolution is evaluated as
\begin{equation}
e^{-i\phi\bar h_p}
=
\hat D(\eta_p)\hat S(\zeta_p)
e^{-i\varpi_p\phi\nh}e^{-i\phi C_p}
\hat S^\dagger(\zeta_p)\hat D^\dagger(\eta_p), 
\end{equation}
and the LP quantum-Volkov operator is thus $\Uh_p(\phi)=e^{i\phi\nh}e^{-i\phi\bar h_p}$.

For a head-on collision geometry, the displacement vanishes, and the quantum-Volkov dressing of an LP pump reduces to a pump-dependent Gaussian transformation,
\begin{align}
\Uh_p(\phi)
&=
e^{i\phi\nh}
\Sh(\zeta_p)
e^{-i(\varpi_p\nh+C_p)\phi}
\Sh^\dagger(\zeta_p),
\label{eq:ULP_headon}
\end{align}
which combines squeezing and number rotation. Its effective squeezing magnitude $r_{\rm eff}(\phi)$ satisfies
\begin{equation}
\sinh r_{\rm eff}(\phi)
=\left|\sinh(2r_p)\sin(\varpi_p\phi)\right|,
\label{eq:LP-effective-squeezing}
\end{equation}
while the squeezing-axis orientation follows from the phases of the associated Bogoliubov coefficients. Thus, even in a strictly head-on collision with no linear momentum exchange along the transverse field direction, the intensity-dependent dressing generates quadrature squeezing.

\begin{figure}[thb]
    \centering
    \includegraphics[width=\columnwidth]{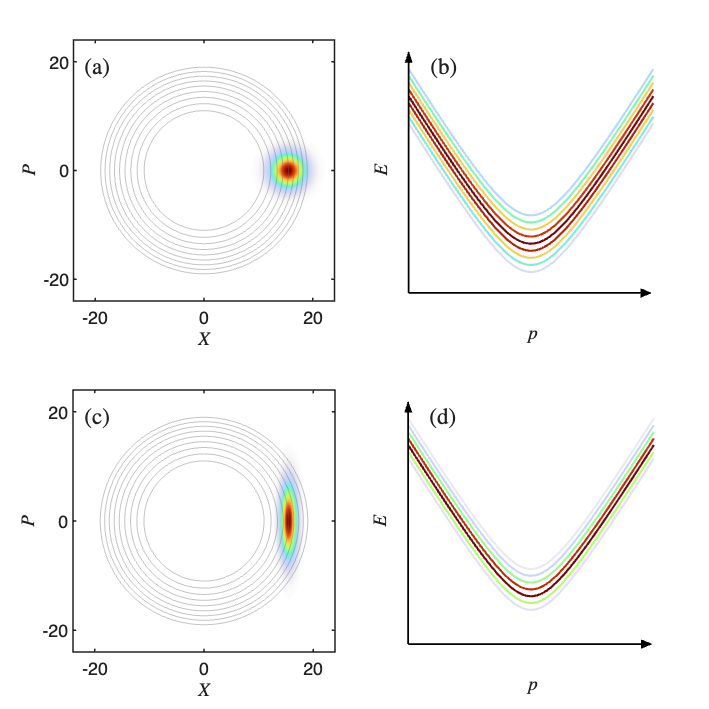}
    \caption{
    Phase-space visualization of quantum-Volkov pump states (left) and the pump-dressed ladder (right).
    The top and bottom rows show circularly and linearly polarized (CP and LP) pump dressing, respectively.
    }
    \label{fig:qvolkov-phase-space}
\end{figure}

Figure~\ref{fig:qvolkov-phase-space}(c) shows the amplitude-squeezed Volkov state generated by a linearly polarized coherent-state pump. The amplitude squeezing reduces the radial variance \ie the photon-number variance of the dressed pump component.

Similar quantum-Volkov states in the nonrelativistic limit were reported by Bergou and Varró~\cite{BergouVarro1981Nonrel, BergouVarro1981Rel, Varro2022}, and quantized-Volkov constructions have also been developed for elliptically polarized fields~\cite{GuoAberg1988}. We nevertheless include the details for completeness and to support the discussion below.

The quantum-Volkov state can be interpreted as a pump-dressed ladder, as shown in Figs.~\ref{fig:qvolkov-phase-space}(b) and \ref{fig:qvolkov-phase-space}(d) for CP and LP, respectively. It is an entangled electron--pump state containing a superposition of correlated components $|p+\ell k,n-\ell\rangle$ for integer $\ell$, in which the electron momentum and pump photon number are constrained by conservation of the total momentum $p^\mu+n k^\mu$.
The phase-space distribution of $\hat U_p|\Psi\rangle_L$ determines the ladder population. The photon-number rungs correspond to radial annuli in the quadrature plane, with $n=(X^2+P^2-1)/2$. The radial extent of the dressed distribution therefore determines the range of populated rungs, while its angular localization determines the optical phase coherence between them. For a bright dressed pump with mean photon number $\bar n_p$, an effective squeeze $r_{\rm eff}$ whose squeezed axis is locally radial gives the leading scaling $(\Delta n_p)^2\simeq\bar n_p e^{-2r_{\rm eff}}$. In that configuration, the LP quantum-Volkov state occupies a narrower range of pump photon-number rungs than the corresponding CP state with the same mean photon number.
This quantum-statistical narrowing is distinct from the classical enhancement associated with LP pump, which results from its larger peak electric field at fixed average intensity.


This ladder picture also clarifies the relation to the classical Volkov state. In the classical description, the pump is replaced by a prescribed $c$-number plane wave, and the electron moves in a periodic background. Because nonlinear Compton harmonics arise from the Fourier components of the Volkov current, the selection rules can be written as $q_i^\mu+N k^\mu=q_f^\mu+k'^\mu$, where $q_i^\mu$ and $q_f^\mu$ are the initial and final Volkov quasi-momenta, $k'^\mu$ is the emitted photon momentum, and $N$ is the harmonic order. In the quantized-pump picture, the same integer has a photon-number interpretation as the net number of pump photons transferred during emission. The classical theory is recovered in the bright, undepleted-pump limit, where many rungs contribute coherently and removing a finite number of pump photons does not appreciably change the pump phase-space distribution.

\section{Scattering amplitude}
\label{sec:amplitude}

We now use the quantum-Volkov states to construct a NLC scattering amplitude that is resolved in the Hilbert space of the driving mode. 
The pump is included nonperturbatively in the Furry Hamiltonian, whereas
emission into all other radiation modes is generated to first order by $H_{\rm rad}=e\int d^3x\,\widehat{\bar\psi}\slashed{\Ah}_{\rm rad}\hat\psi$. 
The perturbative expansion applies only to the emitted radiation field.  The calculation below assumes that the emitted modes are initially in vacuum.  A non-vacuum radiation state can be included without changing the pump dressing by retaining the absorption and stimulated-emission contributions~\cite{DiPiazzaQu2026PRL}.

When positron dynamics are neglected, the electron-sector
interaction separates as
\begin{align}
\hat H_{\rm rad}(t)
&=
\hat H_{\rm emit}(t)+\hat H_{\rm abs}(t),
\label{eq:Hrad_electron_sector} \\
\hat H_{\rm emit}(t)
&=
e\sum_{\sigma,\sigma',\varsigma}
\int
\frac{d^3p\,d^3p'\,d^3k'}{(2\pi)^6}
\frac{\delta^{(3)}(\mathbf p-\mathbf p'-\mathbf k')}
{\sqrt{2E_{\mathbf p}\,2E_{\mathbf p'}\,2\omega'}}
\nonumber\\
&\!\!\!\!\!\!\!\!\times
\bar u_{\sigma'}(p')
\slashed{\epsilon}_{k'\varsigma}^{*}
 u_{\sigma}(p)
\hat c_{\mathbf p'\sigma'}^\dagger
\hat c_{\mathbf p\sigma}
\hat b_{\mathbf k'\varsigma}^\dagger
e^{i(E_{\mathbf p'}+\omega'-E_{\mathbf p})t},
\label{eq:H_emit_cd}
\\
\hat H_{\rm abs}(t)
&=
e\sum_{\sigma,\sigma',\varsigma}
\int
\frac{d^3p\,d^3p'\,d^3k'}{(2\pi)^6}
\frac{\delta^{(3)}(\mathbf p+\mathbf k'-\mathbf p')}
{\sqrt{2E_{\mathbf p}\,2E_{\mathbf p'}\,2\omega'}}
\nonumber\\
&\!\!\!\!\!\!\!\!\times
\bar u_{\sigma'}(p')
\slashed{\epsilon}_{k'\varsigma}
 u_{\sigma}(p)
\hat c_{\mathbf p'\sigma'}^\dagger
\hat c_{\mathbf p\sigma}
\hat b_{\mathbf k'\varsigma}
 e^{i(E_{\mathbf p'}-\omega'-E_{\mathbf p})t}.
\label{eq:H_abs_cd}
\end{align}
Here, $\hat c_{\mathbf p\sigma}$ annihilates the incoming electron, while $\hat b_{\mathbf k'\varsigma}^\dagger$ creates the emitted photon.  The spatial delta function enforces three-momentum conservation at the perturbative radiation vertex. The electron states at this vertex are pump-dressed quantum-Volkov states. Photon emission should be viewed as a transition between two dressed electron--pump ladders, of the type illustrated in Figs.~\ref{fig:qvolkov-phase-space}(b) and \ref{fig:qvolkov-phase-space}(d). 
Each rung corresponds to a correlated component $(p+\ell k,n-\ell)$ of the joint electron--pump state.  
The emission of a photon with momentum $k'$ and polarization $\varsigma$ connects an initial dressed ladder associated with $p$ to a final dressed ladder associated with $p'$, while the ladder rung changes from $n$ to $m$: $\ket{p,\sigma;n}_L\ket{0}_{\rm rad}
    \longrightarrow
    \ket{p',\sigma';m}_L\ket{k',\varsigma}_{\rm rad}$. 
The emitted photon is created by $\hat b_{\mathbf k'\varsigma}^\dagger$, while the
change $n\to m$ of the strong driving mode is contained in the pump-space
matrix element between the initial and final quantum-Volkov dressings.  Thus
NLC radiation is described as a single perturbative emission
event between two nonperturbatively dressed electron--pump states.

An arbitrary quantum state of the pump mode can be expanded in the Fock basis as $\ket{\Psi_i}_L = \sum_n C_n \ket{n}$.
To first order in the perturbative radiation field, the amplitude for
emitting one photon with momentum $k'$ and polarization $\varsigma$ while
the pump--electron dressed state changes from $\ket{p,\sigma;\Psi_i}_L\ket{0}_{\rm rad}$ to $\ket{p',\sigma';\Psi_f}_L\ket{k',\varsigma}_{\rm rad}$ is
\begin{multline}
S_{fi}
=
-\frac{ie}{\sqrt{2p^0\,2p'^0\,2\omega'}}
\int d^4x\,
e^{i(p'+k'-p)\cdot x}
\epsilon_{k'\varsigma,\mu}^{*}
\\
\times
\bar u_{\sigma'}(p')
\mel{\Psi_f}{
\Vh_{p'}^\dagger(\phi)\gamma^\mu\Vh_p(\phi)
}{\Psi_i}
 u_\sigma(p),
\label{eq:Fock-S}
\end{multline}
where
\begin{equation}
\Vh_p(\phi)
=
\left[
1+\frac{e\slashed k\slashed{\Ah}_L(\phi)}{2k\cdot p}
\right]
\Uh_p(\phi).
\label{eq:Vhat}
\end{equation}

We write $k^\mu=\omega\kappa^\mu$, where $\kappa^2=0$, and introduce a
conjugate null vector $\bar\kappa^\mu$ satisfying $\bar\kappa^2=0$ and
$\kappa\cdot\bar\kappa=1$.  The laser phase is $\phi=k\cdot x$.  For any
four-vector $a^\mu$, we define $a_-=\kappa\cdot a$,
$a_+=\bar\kappa\cdot a$, and
$a_\perp^\mu=a^\mu-a_+\kappa^\mu-a_-\bar\kappa^\mu$.  Since the dressed current
depends on spacetime only through $\phi$, the integrations over the three
translationally invariant coordinates give
\begin{align}
p'_-+k'_- -p_-&=0,
\label{eq:minus-conservation}
\\
\mathbf p'_\perp+\mathbf k'_\perp-\mathbf p_\perp&=0.
\label{eq:perp-conservation}
\end{align}
Then the amplitude becomes
\begin{multline}
S_{fi}
=
-\frac{ie(2\pi)^3}{\omega\sqrt{2p^0\,2p'^0\,2\omega'}}
\delta(p'_-+k'_- -p_-)
\delta^{(2)}(\mathbf p'_\perp+\mathbf k'_\perp-\mathbf p_\perp)
\\
\times
\int_{-\infty}^{\infty}d\phi\,
e^{i\lambda_{p'p}^{(e)}\phi}
\bar u_{\sigma'}(p')
\mel{\Psi_f}{
\Uh_{p'}^\dagger(\phi)
\widehat\Gamma_{p'p}^{(\varsigma)}(\phi)
\Uh_p(\phi)
}{\Psi_i}
 u_\sigma(p),
\label{eq:S-phase-integral}
\end{multline}
where
\begin{align}
\lambda_{p'p}^{(e)}
&=
\frac{p'_++k'_+-p_+}{\omega},
\label{eq:lambda-emission} \\
\widehat\Gamma_{p'p}^{(\varsigma)}(\phi)
&=
\left[
1+\frac{e\slashed{\Ah}_L(\phi)\slashed k}{2k\cdot p'}
\right]
\slashed\epsilon_{k'\varsigma}^{*}
\left[
1+\frac{e\slashed k\slashed{\Ah}_L(\phi)}{2k\cdot p}
\right].
\label{eq:Gamma-def}
\end{align}
The delta functions in Eq.~\eqref{eq:S-phase-integral} conserve the
momentum components associated with the directions in which the plane wave is
translationally invariant.  All nontrivial energy exchange with the pump is
left in the integral over $\phi$.  This is the quantum counterpart of the
harmonic Fourier integral in the classical Volkov theory, but here it acts on
the pump Hilbert space.  


The phase integral can be evaluated using a rotating-frame representation of the quantum-Volkov operator. The explicit phase dependence of the pump field is generated by the photon number operator. Using $e^{i\phi\nh}\ah e^{-i\phi\nh}=e^{-i\phi}\ah$ and hence $\Ah_L(\phi)=e^{i\phi\nh}\Ah_L(0)e^{-i\phi\nh}$, we obtain $\widehat\Gamma_{p'p}^{(\varsigma)}(\phi)=e^{i\phi\nh}\widehat\Gamma_{p'p}^{(\varsigma)}(0)e^{-i\phi\nh}$. Using the rotating-frame solution $\Uh_p(\phi)=e^{i\phi\nh}e^{-i\phi\bar h_p}$ from Eq.~\eqref{eq:rotating_hbar_solution}, we obtain
\begin{equation}
\Uh_{p'}^\dagger(\phi)
\widehat\Gamma_{p'p}^{(\varsigma)}(\phi)
\Uh_p(\phi)
=
e^{i\phi\bar h_{p'}}
\widehat\Gamma_{p'p}^{(\varsigma)}(0)
e^{-i\phi\bar h_p}.
\label{eq:vertex-rotating-frame}
\end{equation}
The phase dependence is thus generated by the two phase-independent operators $\bar h_{p'}$ and $\bar h_p$.
This result can be written compactly by introducing the superoperator
\begin{equation}
\mathscr L_{p'p}[X]
=
\bar h_{p'}X-X\bar h_p, 
\label{eq:Liouvillian-definition}
\end{equation}
whose exponential is $ e^{i\phi\mathscr L_{p'p}}[X] = e^{i\phi\bar h_{p'}}X e^{-i\phi\bar h_p}$. 
Using $\int_{-\infty}^{\infty}d\phi\,e^{i\phi X}=2\pi\delta(X)$ in the spectral sense, the phase integral becomes
\begin{multline}
\mathcal I_{fi}
=2\pi\,\bar u_{\sigma'}(p')
\\
\times
\mel{\Psi_f}{
\delta\!\left(
\lambda_{p'p}^{(e)}+\mathscr L_{p'p}
\right)
\widehat\Gamma_{p'p}^{(\varsigma)}(0)
}{\Psi_i}
u_\sigma(p).
\label{eq:I-emission-spectral}
\end{multline}
Projecting the pump states onto the eigenstates, \ie $\bar h_p\ket{n;p}=\varepsilon_n^{(p)}\ket{n;p}$ and $\bar h_{p'}\ket{m;p'}=\varepsilon_m^{(p')}\ket{m;p'}$, gives
\begin{multline}
\mathcal I_{fi}
=
2\pi\sum_{m,n}
\delta\!\left[
\lambda_{p'p}^{(e)}
+\varepsilon_m^{(p')}
-\varepsilon_n^{(p)}
\right]
\bar u_{\sigma'}(p')
\braket{\Psi_f}{m;p'}
\\
\times
\mel{m;p'}{
\widehat\Gamma_{p'p}^{(\varsigma)}(0)
}{n;p}
\braket{n;p}{\Psi_i}
 u_\sigma(p).
\label{eq:I-emission-dressed-basis}
\end{multline}
The delta function obtained from the phase integral enforces
\begin{equation}
p'_++k'_+-p_+
+\omega\left[
\varepsilon_m^{(p')}-\varepsilon_n^{(p)}
\right]
=0.
\label{eq:exact-lightfront-conservation}
\end{equation}
Together with Eqs.~\eqref{eq:minus-conservation} and
\eqref{eq:perp-conservation}, this is the full conservation law for a
transition between the two dressed ladders.  The overlaps
$\braket{n;p}{\Psi_i}$ and $\braket{\Psi_f}{m;p'}$ describe how the
asymptotic pump states populate the incoming and outgoing ladders, while the
middle matrix element is the elementary transition between two ladder rungs illustrated in Figs.~\ref{fig:qvolkov-phase-space}(b)
and \ref{fig:qvolkov-phase-space}(d). The dressed-energy difference between each pair of rungs contributes to the emitted-photon energy. For CP, the two ladders
are displaced number ladders, whereas for LP their relative displacement,
squeezing, and spacing all enter the selection rule.  The usual integer
harmonic condition is recovered only when the occupied rungs are sufficiently
dense and the removal of a finite number of photons does not resolve the
change of the pump.

To evaluate the transition matrix element $\mel{m;p'}{\widehat\Gamma_{p'p}^{(\varsigma)}(0)}{n;p}$ in Eq.~\eqref{eq:Gamma-def}, we expand the vertex into a constant term, terms linear in $\slashed{\Ah}_L$, and a quadratic term proportional to $\slashed{\Ah}_L\slashed{k}\slashed{\epsilon}_{k'\varsigma}^*\slashed{k}\slashed{\Ah}_L$. We choose a gauge in which the emitted photon is transverse to the incident laser direction, $k\cdot\epsilon_{k'\varsigma}^*=0$. The identity $\slashed{k}\slashed{\epsilon}_{k'\varsigma}^*\slashed{k}=2(k\cdot\epsilon_{k'\varsigma}^*)\slashed{k}$ then makes the quadratic term vanish. The remaining vertex contains three terms:
\begin{equation}
\widehat\Gamma_{p'p}^{(\varsigma)}(0)
=
\slashed\epsilon_{k'\varsigma}^{*}
+\frac{e}{2k\cdot p'}
\slashed{\Ah}_L(0)\slashed k
\slashed\epsilon_{k'\varsigma}^{*}
+\frac{e}{2k\cdot p}
\slashed\epsilon_{k'\varsigma}^{*}\slashed k
\slashed{\Ah}_L(0).
\label{eq:Vertex-Linear}
\end{equation}
The vertex is therefore linear in $\ah$ and $\ad$. The nonlinear dynamics is generated by the quantum-Volkov evolution surrounding this
local emission vertex.  

For CP, $\slashed{\Ah}_L(0)=A_0(\slashed\epsilon\ah+ \slashed\epsilon^{*}\ad)$ and 
\begin{equation}
\widehat\Gamma_{p'p}^{(\varsigma)}(0)
=
\Gamma_0^{(\varsigma)}
+\Gamma_1^{(\varsigma)}\ah
+\Gamma_2^{(\varsigma)}\ad,
\label{eq:Gamma-CP-operator}
\end{equation}
with
\begin{align}
\Gamma_0^{(\varsigma)}
&=\slashed\epsilon_{k'\varsigma}^{*},
\label{eq:Gamma0-def}
\\
\Gamma_1^{(\varsigma)}
&=\frac{eA_0}{2}
\left(
\frac{\slashed\epsilon\slashed k
\slashed\epsilon_{k'\varsigma}^{*}}{k\cdot p'}
+\frac{\slashed\epsilon_{k'\varsigma}^{*}\slashed k
\slashed\epsilon}{k\cdot p}
\right),
\label{eq:Gamma1-def}
\\
\Gamma_2^{(\varsigma)}
&=\frac{eA_0}{2}
\left(
\frac{\slashed\epsilon^{*}\slashed k
\slashed\epsilon_{k'\varsigma}^{*}}{k\cdot p'}
+\frac{\slashed\epsilon_{k'\varsigma}^{*}\slashed k
\slashed\epsilon^{*}}{k\cdot p}
\right).
\label{eq:Gamma2-def}
\end{align}
The CP dressed states are displaced Fock states
$\ket{n;p}_{\rm CP}=\Dh^\dagger(\eta_p)\ket n$.  
Using
$\Dh(\eta_{p'})\Dh^\dagger(\eta_p)
=\mathcal P_{p'p}\Dh(\xi_{p'p})$ and
$\Dh(\eta_p)\ah\Dh^\dagger(\eta_p)=\ah-\eta_p$, the matrix element can be written as
\begin{align}
&{}_{\rm CP}\mel{m;p'}{
\widehat\Gamma_{p'p}^{(\varsigma)}(0)
}{n;p}_{\rm CP}
=\mathcal P_{p'p}\Big\{
\Gamma_0^{(\varsigma)}\mathcal D_{mn}(\xi_{p'p})
\nonumber\\
&\quad+
\Gamma_1^{(\varsigma)}
\left[
\sqrt n\,\mathcal D_{m,n-1}(\xi_{p'p})
-\eta_p\mathcal D_{mn}(\xi_{p'p})
\right]
\nonumber\\
&\quad+
\Gamma_2^{(\varsigma)}
\left[
\sqrt{n+1}\,\mathcal D_{m,n+1}(\xi_{p'p})
-\eta_p^{*}\mathcal D_{mn}(\xi_{p'p})
\right]
\Big\},
\label{eq:Gamma-CP-dressed}
\end{align}
where $\mathcal D_{mn}(\xi_{p'p})=\mel m{\Dh(\eta_{p'}-\eta_p)}n$,
and $\mathcal P_{p'p}= \exp[ (\eta_{p'}^{*}\eta_p-\eta_{p'}\eta_p^{*})/2 ]$. 

Thus, every CP rung transition reduces to standard displacement matrix
elements.  In the ladder picture of Fig.~\ref{fig:qvolkov-phase-space}(b), $\mathcal D_{mn}$ is the state overlap between two number ladders displaced by the change of electron momentum.  A transition with a large net photon-number change is therefore controlled by the tail of this overlap rather than by a separate high-order perturbative vertex.

For LP, $\epsilon=\epsilon^{*}$ and $\slashed{\Ah}_L(0)= A_0\slashed\epsilon(\ah+\ad)$. Then Eq.~\eqref{eq:Vertex-Linear} gives
\begin{equation}
\widehat\Gamma_{p'p}^{(\varsigma)}(0)
=
K_0^{(\varsigma)}+K_1^{(\varsigma)}(\ah+\ad),
\label{eq:Gamma-LP-operator}
\end{equation}
where
\begin{align}
K_0^{(\varsigma)}
&=\slashed\epsilon_{k'\varsigma}^{*},
\label{eq:K0-def}
\\
K_1^{(\varsigma)}
&=\frac{eA_0}{2}
\left(
\frac{\slashed\epsilon\slashed k
\slashed\epsilon_{k'\varsigma}^{*}}{k\cdot p'}
+\frac{\slashed\epsilon_{k'\varsigma}^{*}\slashed k
\slashed\epsilon}{k\cdot p}
\right).
\label{eq:K1-def}
\end{align}
For the real, aligned squeezing used in Sec.~\ref{sec:qvolkov}, the LP
dressed states are $\ket{n;p}_{\rm LP}=\Dh(\eta_p)\Sh(r_p)\ket n$.
Define their relative Gaussian overlap
\begin{equation}
\mathcal C_{mn}^{p'p}
=
\mel m{
\Sh^\dagger(r_{p'})\Dh^\dagger(\eta_{p'})
\Dh(\eta_p)\Sh(r_p)
}n.
\label{eq:C-LP-definition}
\end{equation}
On the incoming dressed state,
\begin{equation}
\Sh^\dagger(r_p)\Dh^\dagger(\eta_p)
(\ah+\ad)
\Dh(\eta_p)\Sh(r_p)
=
x_p+\Lambda_p(\ah+\ad),
\label{eq:X-LP-transformed}
\end{equation}
where $x_p=\eta_p+\eta_p^{*}$ and $\Lambda_p=e^{-r_p}$.
Substitution into Eq.~\eqref{eq:Gamma-LP-operator} gives
\begin{multline}
{}_{\rm LP}\mel{m;p'}{
\widehat\Gamma_{p'p}^{(\varsigma)}(0)
}{n;p}_{\rm LP}
=
K_0^{(\varsigma)}\mathcal C_{mn}^{p'p}
\\
+
K_1^{(\varsigma)}
\Bigl[
 x_p\mathcal C_{mn}^{p'p}
+\Lambda_p\sqrt n\,\mathcal C_{m,n-1}^{p'p}
+\Lambda_p\sqrt{n+1}\,\mathcal C_{m,n+1}^{p'p}
\Bigr].
\label{eq:Gamma-LP-dressed}
\end{multline}

For a head-on collision, $\eta_p=\eta_{p'}=0$ and
$\mathcal C_{mj}^{p'p}=\mel m{\Sh(r_p-r_{p'})}j$, so
Eq.~\eqref{eq:Gamma-LP-dressed} reduces directly to squeezed-number-state overlaps.  
Because a squeezing operator connects number states of the same parity, the
$K_0$ part of the head-on LP vertex preserves ladder parity, while the
$K_1\hat X$ part connects the opposite-parity sector through the neighboring
$n\pm1$ rungs.

\subsection{Pure and mixed pump states}

For a pure initial pump state $\ket{\psi_L}=\sum_n C_n\ket n$, the amplitude
conditioned on a specified final pump state $\ket{n'}$ is
\begin{equation}
\mathcal A_{n'}=\sum_n C_n S_{n'n}.
\label{eq:Snprime}
\end{equation}
If the final pump state is not measured, the total probability is
\begin{equation}
P=\sum_{n'}\left|\sum_n C_nS_{n'n}\right|^2.
\label{eq:unobserved-sum}
\end{equation}
The order of the two sums in Eq.~\eqref{eq:unobserved-sum} is physically significant.
Amplitudes originating from different components of the same initial pump state are added before taking the modulus squared, whereas orthogonal, unobserved final pump states are summed incoherently.  Tracing over the final pump therefore does not by itself turn the initial pump into a classical photon-number mixture.

For a general initial pump state $\rhoh_L$, which may be mixed, we define the channel operator
\begin{multline}
\Kh_y
=
-\frac{ie}{\sqrt{2p^0\,2p'^0\,2\omega'}}
\int d^4x\,
e^{i(p'+k'-p)\cdot x}
\epsilon_{k'\varsigma,\mu}^{*}
\\
\times
\bar u_{\sigma'}(p')
\Vh_{p'}^\dagger(\phi)\gamma^\mu
\Vh_p(\phi)
u_\sigma(p).
\label{eq:Kraus}
\end{multline}
where $\Kh_y\equiv\Kh_{p'k'\sigma'\sigma\varsigma}$ and $y\equiv(p',k', \sigma',\sigma,\varsigma)$ denotes the scattering parameters for a fixed initial electron momentum $p$. The corresponding probability density is
\begin{equation}
P_y
=
\Tr_L\!\left(
\Kh_y\rhoh_L\Kh_y^\dagger
\right).
\label{eq:channel-probability}
\end{equation}
This formalism also gives the final pump state conditioned on observing the outcome $y$
\begin{equation}
\rhoh_{L|y}'
=
\frac{\Kh_y\rhoh_L\Kh_y^\dagger}
{\Tr_L\!\left(\Kh_y\rhoh_L\Kh_y^\dagger\right)}.
\label{eq:conditional-pump}
\end{equation}
The corresponding conditional change in the mean pump occupation is
\begin{equation}
\Delta N_{L|y}
=
\Tr_L\!\left(\nh\rhoh_L\right)
-
\Tr_L\!\left(\nh\rhoh_{L|y}'\right).
\label{eq:depletion}
\end{equation}
Thus, the channel operator that determines the emission probability also
determines the measurement back-action on the driving field. In particular,
$\Delta N_{L|y}$ may be conditioned on the emitted-photon momentum and
polarization and on the electron spin, an aspect that can be investigated in future work. 

After summing over the unobserved final electron spin and integrating over final-state phase space, we obtain the differential emission probability
\begin{equation}
dP
=
\sum_{\sigma',\varsigma}
\frac{d^3p'}{(2\pi)^3 2p'^0}
\frac{d^3k'}{(2\pi)^3 2\omega'}
\Tr_L\!\left(\Kh_{p'k'}\rhoh_L\Kh_{p'k'}^\dagger\right).
\label{eq:inclusive-K}
\end{equation}
Equations~\eqref{eq:Kraus}--\eqref{eq:inclusive-K} retain depletion,
back-action, and coherences of the final pump state. They define the exact
operator model used below. Section~\ref{sec:phase-space-evaluation}
recovers the classical-branch average only after imposing the additional
assumption that this back-action is negligible on the scale of the pump
distribution.

\section{Emission spectrum from a Fock-state pump}\label{sec:fock-spectrum}

A Fock state, also called a photon-number state, is considered the most nonclassical quantum state~\cite{Agarwal_2012}. It has a sharp occupation, $\avg{\nh}=n_0$ and $(\Delta n)^2=0$, but no definite optical phase. Although it admits a generalized Glauber--Sudarshan $P$ representation, that representation is singular and cannot be interpreted as a positive statistical mixture of classical amplitudes. The natural description of this state is the dressed-ladder formulation.

In the numerical example, we choose $n_0=100$ so that the discrete pump ladder and its transition matrix elements remain computationally tractable. For an optical field, this occupation would be far below that of a realistic intense pulse. We nevertheless take $a_0\equiv |e|E_0/(m_e\omega)=2$ to examine nonperturbative multiphoton emission, where $E_0$ is the peak pump electric-field amplitude.
To satisfy both conditions, we choose a pump photon energy $\omega=10~\mathrm{keV}$, corresponding (with $\hbar$ restored) to $\lambda\simeq0.124~\mathrm{nm}$, and an initial electron energy of $10~\mathrm{MeV}$, corresponding to $\gamma\simeq19.6$. The emission is integrated over all angles. These parameters define a theoretical benchmark in which moderate occupation and strong coupling coexist, allowing the discrete quantum-Volkov structure to be resolved without the computational cost of a macroscopically occupied mode.

\begin{figure}[htbp]
    \centering
    \includegraphics[width=\columnwidth]{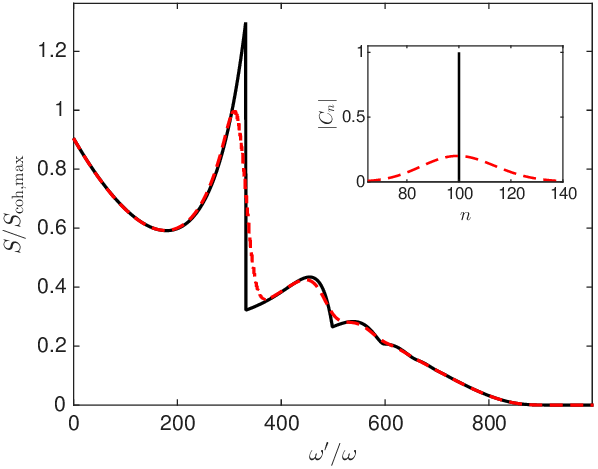}
    \caption{Normalized nonlinear Compton spectra driven by a CP Fock-state pump $\ket{n_0}=\ket{100}$ (black) and a coherent-state pump (red dashed) with mean photon number $\bar n = 100$. The inset shows the photon-number distribution of each pump state.}
    \label{fig:fock_spectrum}
\end{figure}

For a CP pump initially in $\ket{n_0}=\ket{100}$, we evaluate Eq.~\eqref{eq:I-emission-dressed-basis} using the displacement matrix elements in Eq.~\eqref{eq:Gamma-CP-dressed}. For an outgoing dressed-ladder rung $m$, the net rung transfer is $N=n_0-m$, while Eq.~\eqref{eq:exact-lightfront-conservation} determines the corresponding kinematic support. Because the final pump is not measured, probabilities are summed over mutually orthogonal final pump states according to Eq.~\eqref{eq:unobserved-sum}. The resulting spectrum includes pump depletion, electron recoil, and interference within each resolved transition channel and is shown by the black curve in Fig.~\ref{fig:fock_spectrum}. The red dashed curve shows the spectrum for a coherent-state pump with the same mean photon number, $\bar n=\avg{\nh}=100$.

The strong low-frequency contribution is the usual soft-photon enhancement and contains contributions from all $100$ allowed outgoing rungs. At higher frequencies, the net-transfer channels ($N=1, 2, \dots$) terminate at distinct recoil-dependent Compton edges. The first prominent feature occurs near $\omega'/\omega\simeq331$ and is associated with net absorption of one pump photon ($N=1$). Its kinematic support is determined by relativistic momentum conservation [Eq.~\eqref{eq:exact-lightfront-conservation}]. For the stated parameters, the backscattering endpoint ($\theta'=\pi$, measured from the pump direction) has recoil parameter $u_1=[2/(1+a_0^2/2)](\gamma\omega/m_e)(1-\beta\cos\theta')\simeq0.511$. The $N=1$ channel therefore terminates at $(\gamma m_e/\omega)u_1/(1+u_1)\simeq339$; the nearby maximum of the angle-integrated spectral feature need not coincide with this endpoint. The fixed occupation of the Fock pump prevents averaging over different initial occupations and therefore preserves a sharper edge. In contrast, a coherent state has a Poissonian photon-number distribution with $\Delta n=\sqrt{\bar n}=10$; at finite $\avg{\nh}$, the superposition of the slightly displaced edges from its number-state components rounds the harmonic envelope.
For the terminal transfer channel $N=100$, the recoil parameter becomes $u_{100}\simeq51.1$ and the spectral boundary is located at $(\gamma m_e/\omega)u_{100}/(1+u_{100})\simeq982$, beyond which the spectrum vanishes in this monochromatic benchmark.

Finally, consider the macroscopic limit in which the initial photon number becomes very large ($\bar n \to 10^{16}$ and beyond) while the field intensity, and hence $a_0$, remains fixed. The relative photon-number fluctuation of a coherent state, $\Delta n/\bar n=1/\sqrt{\bar n}$, then vanishes asymptotically. Consequently, the broadening associated with the absorption of different photon numbers, which distinguishes coherent and Fock states at low occupation, disappears. Their phase-insensitive inclusive spectra converge to the same plane-wave limit with a sharply defined intensity, although the Fock state still has no definite optical phase.


\section{Undepleted pump limit and Wigner-function representation}
\label{sec:phase-space-evaluation}

The dressed-ladder formulation retains pump depletion and the final pump state, but direct evaluation becomes impractical when the pump contains a macroscopic number of photons. In this regime, the net photon transfer $N$ is typically negligible relative to the mean occupation, $|N|/\bar n\ll1$. The pump then remains effectively unchanged during an emission event, and the discrete ladder can be replaced by a continuous distribution of field amplitudes. Several phase-space representations of such strong quantum pump states are available in this limit. Khalaf and Kaminer~\cite{KhalafKaminer2023}, for example, used the generalized Glauber distribution $P(\alpha,\beta)$, with independent complex variables $\alpha$ and $\beta$, and reduced the inclusive spectrum to an average over the scaled Husimi $Q$ function. Their formulation maps the quantum-light statistics onto classical coherent-field scattering branches.

Here, we instead use the Wigner function, which provides a complete phase-space representation of an arbitrary quantum state. In particular, all single-mode Gaussian states, including coherent, thermal, squeezed-vacuum, and squeezed-coherent states, have regular Gaussian Wigner functions. Non-Gaussian states also possess Wigner representations; for example, a Fock state has a regular, oscillatory Wigner function that directly displays its nonclassical negativity, whereas its ordinary Glauber--Sudarshan $P$ function is highly singular. Although the generalized-$P$ representation of Ref.~\cite{KhalafKaminer2023} remains formally applicable to such states, it generally requires a doubled phase space with independent complex variables $\alpha$ and $\beta$. The Wigner representation therefore provides a particularly direct description of both the Gaussian quantum-light states and non-Gaussian pump states.

For a pump density operator $\rhoh$, the Wigner function is the Fourier transform of the symmetrically ordered characteristic function,
\begin{equation}
W(\alpha,\alpha^*)
=
\frac{1}{\pi^2}\int d^2\lambda\,
\Tr\!\left[
\rhoh e^{\lambda\ad-\lambda^*\ah}
\right]
e^{\lambda^*\alpha-\lambda\alpha^*}.
\label{eq:Wigner-definition}
\end{equation}
It uses the single physical complex amplitude $\alpha$ and expresses the state average as a two-dimensional real integral while retaining the complete Gaussian quantum statistics.
An operator $\widehat{O}(\ah,\ad)$ is correspondingly represented by its Weyl symbol $O_W(\alpha,\alpha^*)$ which, when applied to Eq.~\eqref{eq:channel-probability}, yields
\begin{equation}
P_y
=
\int d^2\alpha\,
K_{y,W}\star W_L\star K_{y,W}^{*}
\simeq
\int d^2\alpha\,
W_L\left|K_{y,W}\right|^2,
\label{eq:Wigner-Moyal-limit}
\end{equation}
where $\star$ denotes the Moyal product and the second expression is the leading term of the bright, weakly depleted expansion. Accordingly, the emission probability [cf. Eq.~\eqref{eq:inclusive-K}] obtained after tracing over the unobserved final pump states is
\begin{multline}
dP \simeq
\sum_{\sigma',\varsigma}
\frac{d^3p'}{(2\pi)^3 2p'^0}
\frac{d^3k'}{(2\pi)^3 2\omega'}
\\
\times
\int d^2\alpha\,
W_L(\alpha,\alpha^*)
\left|K_{p'k',W}(\alpha,\alpha^*)\right|^2, 
\label{eq:Wigner-Kraus-undepleted}
\end{multline}
where $W_L$ is the Wigner function of the initial pump state and $K_{p'k',W}$ is the emission amplitude for an undepleted branch with field amplitude $\alpha$. In the ladder picture of
Figs.~\ref{fig:qvolkov-phase-space}(b) and \ref{fig:qvolkov-phase-space}(d), this approximation describes transitions within clusters of nearby, macroscopically occupied rungs. It does not resolve the final pump occupation or retain the conditional back-action of the emission event.

At leading order in the bright-pump (Moyal-gradient) expansion, the Wigner--Weyl transform of Eq.~\eqref{eq:Kraus} replaces $\ah$ with $\alpha$ and $\ad$ with $\alpha^*$:
\begin{multline}
K_{p'k',W}(\alpha,\alpha^*)
=
-\frac{ie}{\sqrt{2p^0\,2p'^0\,2\omega'}}
\\
\times
\int d^4x\,e^{i(p'+k'-p)\cdot x}
\bar u_{\sigma'}(p')
\widetilde F_{p'p}(\phi;\alpha)
\Gamma_{p'p,W}^{(\varsigma)}(\phi;\alpha)
u_\sigma(p),
\label{eq:Weyl-Integrand-Map}
\end{multline}
The scalar
factor $\widetilde F_{p'p}$ contains the accumulated Volkov
phase, and
$\Gamma_{p'p,W}^{(\varsigma)}$ is the local spinor vertex.
As shown in Eq.~\eqref{eq:S-phase-integral}, three of the spacetime
integrations produce
$\delta(p'_-+k'_- -p_-)\delta^{(2)}
(\mathbf p'_\perp+\mathbf k'_\perp-\mathbf p_\perp)$.
All nontrivial pump-energy exchange is contained in the remaining
integral over $\phi=k\cdot x$. The cycle average of the branch generator and the corresponding quasimomentum are
\begin{align}
h_{p,0}(\alpha)
&=\frac{1}{2\pi}\int_0^{2\pi}d\phi\,h_p(\phi;\alpha)
\simeq\Omega_p|\alpha|^2,
\label{eq:branch-quasimomentum}
\\
q_p^\mu(\alpha)&=p^\mu+h_{p,0}(\alpha)k^\mu,
\end{align}
where the omitted Weyl-ordering correction is subleading at large occupation. Absorbing the secular phase into $q_p^\mu(\alpha)$ and expanding the remaining periodic integrand in harmonics gives
\begin{multline}
S_{fi}(\alpha)
=-ie(2\pi)^4
\sum_{\ell=-\infty}^{\infty}
\frac{
\delta^{(4)}\!\left[
q_p(\alpha)+\ell k-q_{p'}(\alpha)-k'
\right]
}{
\sqrt{2p^0\,2p'^0\,2\omega'}
}
\\
\times\bar u_{\sigma'}(p')\mathcal M_\ell(\alpha)u_\sigma(p),
\label{eq:S-branch-harmonic}
\end{multline}
\begin{equation}
\mathcal M_\ell(\alpha)
=
\frac{1}{2\pi}
\int_0^{2\pi}d\phi\,
e^{i\ell\phi}
F_{p'p}(\phi;\alpha)
\Gamma_{p'p,W}^{(\varsigma)}
(\phi;\alpha).
\label{eq:Mell-general-definition}
\end{equation}

For CP, $F_{p'p}(\phi;\alpha)$ can be obtained by expressing the associated Laguerre polynomial in Eq.~\eqref{eq:Gamma-CP-dressed} through the asymptotic relation $\mathcal D_{n+\ell,n}(\xi_{p'p}) \to e^{i\ell\delta}J_\ell(z)$ as $n\to\infty$. However, this procedure does not extend directly to the LP pump because its dressing also contains a squeezing operator. A unified expression for both polarizations follows instead by taking the Wigner--Weyl transform of $\Uh_{p'}^\dagger(\phi)\Uh_p(\phi)$ in the phase integral:
\begin{multline}
F_{p'p}(\phi;\alpha)
=
\exp\!\bigg\{-i\int_0^\phi d\varphi\,
\Big[
 h_p(\varphi;\alpha)
-h_{p'}(\varphi;\alpha)
\\
-h_{p,0}(\alpha)
+h_{p',0}(\alpha)
\Big]\bigg\},
\label{eq:periodic-Volkov-factor}
\end{multline}
where $h_{p,0}(\alpha)$ is defined in Eq.~\eqref{eq:branch-quasimomentum}.

\subsection{Circularly polarized pump}

For a CP pump, Eq.~\eqref{eq:hp} gives
\begin{equation}
h_p^{({\rm CP})}(\phi;\alpha)
=
\Omega_p|\alpha|^2
+g_p\alpha e^{-i\phi}
+g_p^{*}\alpha^{*}e^{i\phi}.
\label{eq:h-CP-branch}
\end{equation}
Defining $\beta_{p'p}
=(g_p-g_{p'})\alpha
=|\beta_{p'p}|e^{i\delta_{p'p}}$, and $z_{p'p}=2|\beta_{p'p}|$, we obtain
\begin{align}
F_{p'p}^{({\rm CP})}(\phi;\alpha)
&=
\exp\!\left[
\beta_{p'p}e^{-i\phi}
-\beta_{p'p}^{*}e^{i\phi}
\right] \nonumber \\
&=
\exp\!\left[-iz_{p'p}\sin(\phi-\delta_{p'p})\right].
\label{eq:F-CP-explicit}
\end{align}
The corresponding harmonic coefficients are
\begin{align}
\mathcal I_\ell^{({\rm CP})}(\alpha)
&=
\frac{1}{2\pi}\int_0^{2\pi}d\phi\,
e^{i\ell\phi}F_{p'p}(\phi;\alpha)\nonumber\\
&=
e^{i\ell\delta_{p'p}}J_\ell(z_{p'p}).
\label{eq:Iell-CP}
\end{align}
The CP vertex follows directly from the operator form in Eq.~\eqref{eq:Gamma-CP-operator}:
\begin{equation}
\Gamma_{p'p,W}^{(\varsigma,{\rm CP})}(\phi;\alpha)
=
\Gamma_0^{(\varsigma)}
+\Gamma_1^{(\varsigma)}\alpha e^{-i\phi}
+\Gamma_2^{(\varsigma)}\alpha^{*}e^{i\phi},
\label{eq:Gamma-CP-phase-Weyl}
\end{equation}
Substitution into Eq.~\eqref{eq:Mell-general-definition} gives
\begin{equation}
\mathcal M_\ell^{({\rm CP})}(\alpha)
=
\Gamma_0^{(\varsigma)}\mathcal I_\ell^{({\rm CP})}
+\Gamma_1^{(\varsigma)}\alpha\mathcal I_{\ell-1}^{({\rm CP})}
+\Gamma_2^{(\varsigma)}\alpha^{*}\mathcal I_{\ell+1}^{({\rm CP})}.
\label{eq:Mell-CP}
\end{equation}
The neighboring Bessel coefficients are generated by the explicit factors $e^{\mp i\phi}$ multiplying the amplitude and its conjugate in the local vertex. The harmonic order $\ell$ is the phase-space counterpart of the discrete net ladder separation $N$ identified in Sec.~\ref{sec:fock-spectrum}.

\subsection{Linearly polarized pump}

For an LP pump, Eq.~\eqref{eq:hp} gives
\begin{multline}
h_p^{({\rm LP})}(\phi;\alpha)
=
\Omega_p|\alpha|^2
+g_p\alpha e^{-i\phi}
+g_p^{*}\alpha^{*}e^{i\phi}
\\
+\frac{\Omega_p}{2}
\left(
\alpha^2e^{-2i\phi}
+\alpha^{*2}e^{2i\phi}
\right).
\label{eq:h-LP-branch}
\end{multline}
The first-harmonic part is again characterized by $\beta_{p'p}=(g_p-g_{p'})\alpha$, while the oscillatory $A^2$ term introduces $\gamma_{p'p}
=
\frac{\Omega_p-\Omega_{p'}}{4}\alpha^2
=|\gamma_{p'p}|e^{i\delta_{2,p'p}}$. 
Writing $\beta_{p'p}=|\beta_{p'p}|e^{i\delta_{1,p'p}}$, the periodic phase is
\begin{multline}
F_{p'p}^{({\rm LP})}(\phi;\alpha)
=
\exp\Big[
\beta_{p'p}e^{-i\phi}-\beta_{p'p}^{*}e^{i\phi}
\\
+\gamma_{p'p}e^{-2i\phi}-\gamma_{p'p}^{*}e^{2i\phi}
\Big].
\label{eq:F-LP-explicit}
\end{multline}
Let $z_1=2|\beta_{p'p}|$, $z_2=2|\gamma_{p'p}|$, and $\vartheta_{p'p}=\delta_{2,p'p}-2\delta_{1,p'p}$. Expanding the two periodic factors separately and collecting the coefficient of $e^{-i\ell\phi}$ gives
\begin{equation}
\mathcal I_\ell^{({\rm LP})}(\alpha)
=
e^{i\ell\delta_{1,p'p}}
\sum_{s=-\infty}^{\infty}
J_{\ell-2s}(z_1)J_s(z_2)
e^{is\vartheta_{p'p}},
\label{eq:Iell-LP-generalized-Bessel}
\end{equation}
which is the phase-dependent two-argument generalized-Bessel coefficient. The phase factor can be omitted only in the special case $\vartheta_{p'p}=0$ modulo $2\pi$.

The LP vertex is
\begin{equation}
\Gamma_{p'p,W}^{(\varsigma,{\rm LP})}(\phi;\alpha)
=
K_0^{(\varsigma)}
+K_1^{(\varsigma)}
\left(\alpha e^{-i\phi}+\alpha^{*}e^{i\phi}\right),
\label{eq:Gamma-LP-phase-Weyl}
\end{equation}
and thus,
\begin{equation}
\mathcal M_\ell^{({\rm LP})}(\alpha)
=
K_0^{(\varsigma)}\mathcal I_\ell^{({\rm LP})}
+K_1^{(\varsigma)}
\left[
\alpha\mathcal I_{\ell-1}^{({\rm LP})}
+\alpha^{*}\mathcal I_{\ell+1}^{({\rm LP})}
\right].
\label{eq:Mell-LP}
\end{equation}

For either polarization, Eqs.~\eqref{eq:Mell-CP} and \eqref{eq:Mell-LP} give the branch-averaged differential emission probability
\begin{equation}
\frac{dP^{(\ell)}}{d\Pi_{p'}d\Pi_{k'}}
\simeq
\int d^2\alpha\,W_L(\alpha,\alpha^*)
\frac{dP_{\rm cl}^{(\ell)}(\alpha)}{d\Pi_{p'}d\Pi_{k'}},
\label{eq:dP-phase-space-undepleted}
\end{equation}
where $dP_{\rm cl}^{(\ell)}(\alpha)$ is computed from the spinor matrix element $\bar u_{\sigma'}(p')\mathcal M_\ell(\alpha)u_\sigma(p)$ together with the four-momentum constraint in Eq.~\eqref{eq:S-branch-harmonic}, and $d\Pi_{p'}=d^3p'/(2\pi)^3$, $d\Pi_{k'}=d^3k'/(2\pi)^3$. This form avoids treating the branch-dependent four-dimensional delta function as if it were independent of $\alpha$.
After the final-electron variables are eliminated and the unobserved spin and
polarization states are summed, we denote the frequency spectrum by
$dP/d\omega'$ when it is integrated over the emitted-photon direction and by
$dP/(d\omega' d\Omega')$ when that direction is retained.
Equation~\eqref{eq:dP-phase-space-undepleted} is the continuous, macroscopic limit of the exact operator model. It replaces the computationally intensive discrete dressed-state sums by a two-dimensional phase-space average while retaining the full nonlinear dependence of every undepleted branch.
Equations~\eqref{eq:S-branch-harmonic},
\eqref{eq:Mell-general-definition}, and
\eqref{eq:dP-phase-space-undepleted} summarize the undepleted-pump formulation:
the first gives the harmonic scattering amplitude for a fixed phase-space
branch, the second defines the corresponding harmonic vertex, and the third
gives the differential emission probability after averaging over the Wigner
distribution of the pump.

This result has the same weighted-sum structure as the spectrum reported in Ref.~\cite{KhalafKaminer2023}, with $W_L(\alpha,\alpha^*)$ playing the role of the state weight represented there by $\widetilde Q(\mathcal E_\alpha)$. 
For a macroscopic coherent pump, the absolute Wigner width remains at the vacuum scale, but its width relative to the mean amplitude decreases as $1/|\alpha_0|$. Thus, in a field variable rescaled by the mean amplitude, both the Wigner and Husimi weights localize at the mean field and recover the conventional undepleted classical spectrum. The exact dressed-ladder theory additionally retains the outgoing rung $m$ in the transition weight and kinematics. By tracing over the final pump, the Wigner-function approach removes interference between orthogonal final pump states.

\section{Squeezed coherent driving states}
\label{sec:scs_pump}

The Fock-state example isolates the consequences of a sharply defined pump
occupation, but preparing a high-occupation number state is experimentally
demanding. Squeezed coherent states provide a continuous and more accessible
family in which the mean field and its fluctuations can be varied
independently.  In particular, amplitude squeezing can produce
sub-Poissonian photon statistics and reduce the range of initial dressed
rungs populated by the pump.  It does not make the pump a number state, but
it permits a controlled interpolation between a narrow intensity
distribution and a broad, super-Poissonian one at fixed mean occupation.

This application also illustrates the division between the Wigner function approach and the dressed ladder method.  The squeezed coherent state itself is represented exactly by a smooth Gaussian Wigner function, while the scattering kernel is evaluated in the undepleted approximation of Eq.~\eqref{eq:dP-phase-space-undepleted}.  If final-pump resolution or depletion is required, the same input state can instead be expanded in the Fock basis and inserted into Eq.~\eqref{eq:I-emission-dressed-basis}.  The phase-space calculation is therefore a controlled reduction of the operator model, not a separate physical model.


We define the squeezed coherent pump by
$\ket{\Psi_L}=\Dh(\alpha_0)\Sh(\xi)\ket{0}$, with
$\xi=re^{i\theta}$.  Its Wigner function is a positive Gaussian
\begin{multline}
W_{\rm SCS}(\alpha, \alpha^*) = \frac{2}{\pi} \exp\bigg[ -2 \Big| (\alpha-\alpha_0)\cosh r \\
+ (\alpha^*-\alpha_0^*)e^{i\theta}\sinh r \Big|^2 \bigg].
\label{eq:scs_wigner_function}
\end{multline}
For $r\to0$ this becomes the circular Gaussian of a coherent state centered
at $\alpha_0$, whereas $\alpha_0=0$ gives squeezed vacuum.  The orientation
of the ellipse is set by $\theta/2$; its angle relative to the coherent
displacement will determine the radial, and hence intensity, fluctuations.

We illustrate the effect of squeezing in NLC scattering using numerical simulations of a CP pump with photon energy $\omega=10~\mathrm{keV}$, corresponding (with $\hbar$ restored) to $\lambda\simeq1.24~\text{\AA}$, nonlinearity parameter $a_0=2$, and mean occupation $\bar n=\avg{\nh}=100$. The squeezed states have $8~\mathrm{dB}$ of quadrature squeezing, corresponding to $r=(8/20)\ln 10\simeq0.921$. This idealized finite-occupation parameter set is chosen to make the statistical differences visible and should not be interpreted as a complete source specification, since relating $\bar n$ to $a_0$ also requires the spatial and temporal mode normalization.

\begin{figure}[th]
\centering
\includegraphics[width=\columnwidth]{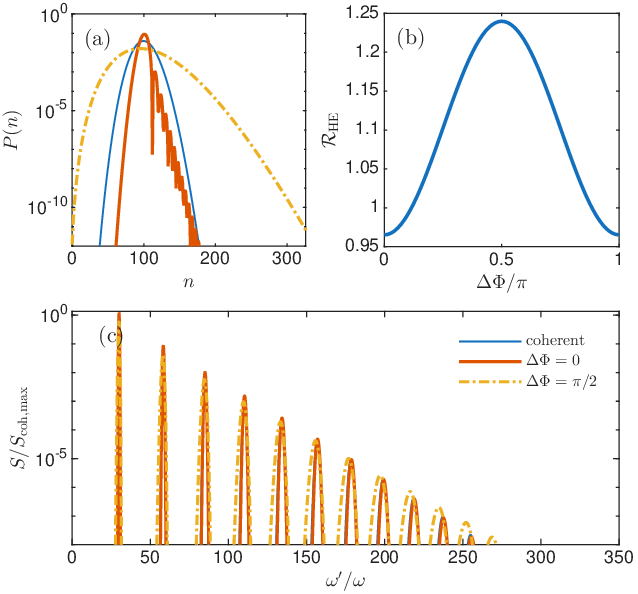}
\caption{(a) Photon-number distributions $P_n=|C_n|^2$ for different pump states.  (b) Relative high-energy
yield $\mathcal R_{\rm HE}$, integrated over $\omega'/\omega>100$, versus the
relative squeezing angle $\Delta\Phi$.  (c) Normalized emission spectra per
solid angle.
}
\label{fig:scs_control}
\end{figure}

To reveal the different photon statistics, we expand the SCS in the Fock-state basis, \ie 
$\ket{\Psi_L}=\sum_{n=0}^{\infty}C_n^{(\alpha_0,\xi)}\ket n$ with 
\begin{multline}
C_n^{(\alpha_0,\xi)}
=
\frac{
\exp\left[
-\frac{1}{2}|\alpha_0|^2
-\frac{1}{2}\zeta \alpha_0^{*2}
\right]
}{\sqrt{n!\cosh r}}
\left(
\frac{\zeta}{2}
\right)^{\frac{n}{2}}
\\
\times
H_n
\left(
\frac{\alpha_0+\zeta\alpha_0^*}{\sqrt{2\zeta}}
\right),
\label{eq:scs_fock_coefficients}
\end{multline}
where $\zeta=e^{i\theta}\tanh r$ and $H_n$ is the Hermite polynomial.
In the limit $r\to0$, this reduces to
$C_n=e^{-|\alpha_0|^2/2}\alpha_0^n/\sqrt{n!}$.  The mean occupation is
$\bar n=|\alpha_0|^2+\sinh^2r$, and the variance is
\begin{multline}
(\Delta n)^2
=
|\alpha_0|^2
\left[
\cosh(2r)
-
\sinh(2r)
\cos(2\phi_0-\theta)
\right] \\
+
2\sinh^2 r\left(1+\sinh^2 r\right),
\label{eq:scs_number_variance}
\end{multline}
where $\alpha_0=|\alpha_0|e^{i\phi_0}$.  The relative angle
$\Delta\Phi=\theta/2-\phi_0$ determines whether the displacement lies along
the squeezed or anti-squeezed quadrature.  In the bright limit
$|\alpha_0|^2\gg\sinh^2r$, Eq.~\eqref{eq:scs_number_variance} becomes
\begin{equation}
(\Delta n)^2
\simeq
|\alpha_0|^2
\left(
e^{-2r}\cos^2\Delta\Phi
+
e^{2r}\sin^2\Delta\Phi
\right).
\label{eq:scs_bright_variance}
\end{equation}
As shown in Fig.~\ref{fig:scs_control}(a), $\Delta\Phi=0$ suppresses photon number fluctuations, whereas
$\Delta\Phi=\pi/2$ aligns the displacement with the anti-squeezed
quadrature and enhances them.  The oscillations in the amplitude-squeezed distribution are the
Schleich--Wheeler oscillations~\cite{Schleich1987}, which originate from
interference between the coherent displacement and the pairwise photon
amplitudes generated by squeezing.


The emission spectra per solid angle are calculated using
Eq.~\eqref{eq:dP-phase-space-undepleted}, and are plotted in Fig.~\ref{fig:scs_control}(c). For direct comparison with Ref.~\cite{KhalafKaminer2023}, the electron counterpropagates with the pump and the emitted-photon polar angle is fixed at $\theta'=159.9^\circ$, in the notation of that reference. This direction is approximately $20.1^\circ$ from the incident-electron axis. The coherent-state pump provides the reference harmonic envelope. Phase squeezing gives greater weight to branches with
large $|\alpha|$, which have larger ponderomotive shifts and support higher
effective harmonic orders; its spectrum is consequently broader and its
high-energy tail decays more slowly.  Amplitude squeezing removes many of
these large-intensity excursions, narrows the range of quasimomenta sampled
in Eq.~\eqref{eq:dP-phase-space-undepleted}, and suppresses the high-energy tail. 

To quantify the tail, we define $\mathcal R_{\rm HE}$ as the ratio of the number of high-energy photons with $\omega'/\omega>100$ driven by an SCS to that driven by a coherent-state pump. 
Figure~\ref{fig:scs_control}(b) shows this spectral redistribution as a function of the squeezing angle $\Delta\Phi$.  The coherent-state result is the reference $\mathcal R_{\rm HE}=1$.  At $\Delta\Phi=0$ and $\pi$, the coherent displacement lies along the squeezed quadrature, causing amplitude squeezing.  The resulting sub-Poissonian number distribution suppresses rare large-$|\alpha|$ branches, so fewer higher energy photons ($\omega'/\omega>100$) are emitted and $\mathcal R_{\rm HE}<1$.  The two endpoint minima are equivalent because a squeezing ellipse is unchanged by a rotation through $\pi$.  Conversely, the maximum near $\Delta\Phi=\pi/2$ occurs when the displacement lies along the anti-squeezed quadrature, leading to phase squeezing.  The enhanced photon number fluctuations then populate a long high-energy tail.  Those relatively rare branches support higher effective harmonic orders and contribute disproportionately to the high-energy integral, giving $\mathcal R_{\rm HE}>1$.

This quantum-statistical effect should be distinguished from the semiclassical squeezed-background dynamics studied in Ref.~\cite{DiPiazzaQu2026FM}.  There, quantum fluctuations of the prepared state were neglected and squeezing primarily produced an effective frequency modulation of the prescribed classical field.  Here the mean waveform  is fixed and the Wigner distribution of field amplitudes is retained, so the observable changes because different intensity fluctuations sample the nonlinear Compton kernel with different weights.


For LP there is an additional source of squeezing arising from the
quadratic $A^2$ term in the quantum-Volkov generator.  The exact matrix
element then contains the externally prepared state acted on by
$\Uh_p^{\rm LP}(\phi)$, and depends on the relative orientation of the input
squeezing and the dressing-induced squeezing.  Depending on this orientation,
the two transformations can increase or partially cancel the radial
variance.  In the exact ladder language this behavior is carried by the
relative Gaussian overlaps $\mathcal C_{mn}^{p'p}$; in the undepleted
Wigner representation it is carried by the two-argument generalized Bessel coefficient
$\mathcal I_\ell^{({\rm LP})}$.

\section{Conclusions}
\label{sec:conclusion}

In conclusion, we have developed a fully quantized, single-mode Furry-picture formulation of nonlinear Compton scattering.  The driving mode is incorporated nonperturbatively into the electron dressing, while the emitted radiation is treated to first order.  The resulting amplitudes resolve both the initial and final pump states and therefore retain pump depletion, back-action, and the correlations generated among the pump, electron, and emitted photon. Final-pump-resolved observables follow directly, whereas the usual inclusive spectrum is obtained by tracing over the unobserved final pump.
The quantum-Volkov evolution has a simple polarization-dependent Gaussian structure.  For CP, it is a displacement followed by a number rotation; for LP, the quadratic pump-field term additionally produces squeezing.  The Fock-state example illustrates that the spectrum resolves recoil-dependent harmonic structures and terminates at a finite-energy boundary fixed by the bounded pump photon occupation.  These features are obscured when the pump is replaced by a coherent field.

In the undepleted limit, we developed a Wigner--Weyl representation of the pump quantum state and established its connection to the exact dressed-ladder theory.  Starting from the inclusive sum over final pump states, this phase-space representation reduces to an average of nonlinear Compton probabilities over components with different amplitudes, similar to the generalized $P$-function approach reported in Ref.~\cite{KhalafKaminer2023}. Unlike the formulation in Ref.~\cite{KhalafKaminer2023}, the Wigner-function representation provides a complex phase-space description that is regular for Gaussian quantum states and avoids the doubled complex phase-space integral. The phase integral gives ordinary Bessel coefficients for CP and generalized Bessel coefficients for LP.  
The squeezed coherent state example shows that the emission is governed by both the mean pump occupation and its fluctuations.  At a fixed mean occupation, amplitude squeezing suppresses high-photon-number branches and the high-energy emission, whereas phase squeezing enhances the tail by increasing the weight of such branches.  

The sharp edges of the harmonic structure rely on the monochromatic single-mode approximation. In a finite pulse, the spectrum becomes continuous and develops duration-dependent broadening and substructure; in the few-cycle regime, the usual monochromatic effective-mass signature can also be modified or absent~\cite{BocaFlorescu2009,SeiptKampfer2011,MackenrothDiPiazza2011}. The present framework can be extended to pulsed and multimode pumps, repeated emission, and conditional measurements of the outgoing light. Such extensions  will be required to determine when pump depletion and quantum correlations become observable in realistic strong-field experiments and when an inclusive phase-space description remains sufficient.

From an experimental perspective, the exact quantum treatment is most relevant when the interaction transfers a non-negligible fraction of the pump occupation or resolves photon-number-dependent correlations. At a fixed intensity, the macroscopic limit requires the single-photon field amplitude to scale as $A_0\propto\bar n^{-1/2}$. For a coherent state, $\Delta n/\bar n=1/\sqrt{\bar n}\to0$, while a Fock state has $\Delta n=0$ identically; consequently, their phase-insensitive inclusive spectra converge whenever the transferred photon number satisfies $N\ll n_0$. The discrete transition ladder then becomes effectively continuous and the standard classical Volkov theory is recovered. Observable departures from this limit require either $N/n_0$ to be appreciable or nonclassical fluctuations and coherences to remain resolvable. Their practical realization therefore favors a strongly coupled mode with moderate occupation, achieved, for example, through high photon energy, tight confinement, or a small effective mode volume, together with sufficient spectral resolution or joint measurement of the emitted radiation and residual pump. For conventional high-intensity optical pulses, whose relevant modes are macroscopically occupied and only weakly depleted, these conditions are generally not met, and the semiclassical description remains an excellent approximation.

\begin{acknowledgments}
The author expresses gratitude to Prof. Antonino Di Piazza for his invaluable discussions and to Prof. Nathaniel J. Fisch for providing the research environment.
This work was supported by NNSA Grant No. DE-NA0004167, NSF Grant No. PHY-2308829, and
the Center for Magnetic Acceleration, Compression, and Heating (MACH), part of the U.S. DOE-NNSA Stewardship Science Academic Alliances Program under Cooperative Agreement No. DE-NA0004148.
\end{acknowledgments}

\bibliography{NCS_refs}

\begin{thebibliography}{36}%
\makeatletter
\providecommand \@ifxundefined [1]{%
 \@ifx{#1\undefined}
}%
\providecommand \@ifnum [1]{%
 \ifnum #1\expandafter \@firstoftwo
 \else \expandafter \@secondoftwo
 \fi
}%
\providecommand \@ifx [1]{%
 \ifx #1\expandafter \@firstoftwo
 \else \expandafter \@secondoftwo
 \fi
}%
\providecommand \natexlab [1]{#1}%
\providecommand \enquote  [1]{``#1''}%
\providecommand \bibnamefont  [1]{#1}%
\providecommand \bibfnamefont [1]{#1}%
\providecommand \citenamefont [1]{#1}%
\providecommand \href@noop [0]{\@secondoftwo}%
\providecommand \href [0]{\begingroup \@sanitize@url \@href}%
\providecommand \@href[1]{\@@startlink{#1}\@@href}%
\providecommand \@@href[1]{\endgroup#1\@@endlink}%
\providecommand \@sanitize@url [0]{\catcode `\\12\catcode `\$12\catcode
  `\&12\catcode `\#12\catcode `\^12\catcode `\_12\catcode `\%12\relax}%
\providecommand \@@startlink[1]{}%
\providecommand \@@endlink[0]{}%
\providecommand \url  [0]{\begingroup\@sanitize@url \@url }%
\providecommand \@url [1]{\endgroup\@href {#1}{\urlprefix }}%
\providecommand \urlprefix  [0]{URL }%
\providecommand \Eprint [0]{\href }%
\providecommand \doibase [0]{https://doi.org/}%
\providecommand \selectlanguage [0]{\@gobble}%
\providecommand \bibinfo  [0]{\@secondoftwo}%
\providecommand \bibfield  [0]{\@secondoftwo}%
\providecommand \translation [1]{[#1]}%
\providecommand \BibitemOpen [0]{}%
\providecommand \bibitemStop [0]{}%
\providecommand \bibitemNoStop [0]{.\EOS\space}%
\providecommand \EOS [0]{\spacefactor3000\relax}%
\providecommand \BibitemShut  [1]{\csname bibitem#1\endcsname}%
\let\auto@bib@innerbib\@empty
\bibitem [{\citenamefont {Volkov}(1935)}]{Volkov1935}%
  \BibitemOpen
  \bibfield  {author} {\bibinfo {author} {\bibfnamefont {D.~M.}\ \bibnamefont
  {Volkov}},\ }\bibfield  {title} {\bibinfo {title} {{\"U}ber eine klasse von
  l{\"o}sungen der {Diracschen} gleichung},\ }\href
  {https://doi.org/10.1007/BF01331022} {\bibfield  {journal} {\bibinfo
  {journal} {Z. Phys.}\ }\textbf {\bibinfo {volume} {94}},\ \bibinfo {pages}
  {250} (\bibinfo {year} {1935})}\BibitemShut {NoStop}%
\bibitem [{\citenamefont {Brown}\ and\ \citenamefont
  {Kibble}(1964)}]{BrownKibble1964}%
  \BibitemOpen
  \bibfield  {author} {\bibinfo {author} {\bibfnamefont {L.~S.}\ \bibnamefont
  {Brown}}\ and\ \bibinfo {author} {\bibfnamefont {T.~W.~B.}\ \bibnamefont
  {Kibble}},\ }\bibfield  {title} {\bibinfo {title} {Interaction of intense
  laser beams with electrons},\ }\href
  {https://doi.org/10.1103/PhysRev.133.A705} {\bibfield  {journal} {\bibinfo
  {journal} {Phys. Rev.}\ }\textbf {\bibinfo {volume} {133}},\ \bibinfo {pages}
  {A705} (\bibinfo {year} {1964})}\BibitemShut {NoStop}%
\bibitem [{\citenamefont {Nikishov}\ and\ \citenamefont
  {Ritus}(1964)}]{NikishovRitus1964}%
  \BibitemOpen
  \bibfield  {author} {\bibinfo {author} {\bibfnamefont {A.~I.}\ \bibnamefont
  {Nikishov}}\ and\ \bibinfo {author} {\bibfnamefont {V.~I.}\ \bibnamefont
  {Ritus}},\ }\bibfield  {title} {\bibinfo {title} {Quantum processes in the
  field of a plane electromagnetic wave and in a constant field},\ }\href@noop
  {} {\bibfield  {journal} {\bibinfo  {journal} {Sov. Phys. JETP}\ }\textbf
  {\bibinfo {volume} {19}},\ \bibinfo {pages} {1191} (\bibinfo {year}
  {1964})}\BibitemShut {NoStop}%
\bibitem [{\citenamefont {Ritus}(1985)}]{Ritus1985}%
  \BibitemOpen
  \bibfield  {author} {\bibinfo {author} {\bibfnamefont {V.~I.}\ \bibnamefont
  {Ritus}},\ }\bibfield  {title} {\bibinfo {title} {Quantum effects of the
  interaction of elementary particles with an intense electromagnetic field},\
  }\href {https://doi.org/10.1007/BF01120220} {\bibfield  {journal} {\bibinfo
  {journal} {J. Sov. Laser Res.}\ }\textbf {\bibinfo {volume} {6}},\ \bibinfo
  {pages} {497} (\bibinfo {year} {1985})}\BibitemShut {NoStop}%
\bibitem [{\citenamefont {Di~Piazza}\ \emph {et~al.}(2012)\citenamefont
  {Di~Piazza}, \citenamefont {M{\"u}ller}, \citenamefont {Hatsagortsyan},\ and\
  \citenamefont {Keitel}}]{DiPiazza2012}%
  \BibitemOpen
  \bibfield  {author} {\bibinfo {author} {\bibfnamefont {A.}~\bibnamefont
  {Di~Piazza}}, \bibinfo {author} {\bibfnamefont {C.}~\bibnamefont
  {M{\"u}ller}}, \bibinfo {author} {\bibfnamefont {K.~Z.}\ \bibnamefont
  {Hatsagortsyan}},\ and\ \bibinfo {author} {\bibfnamefont {C.~H.}\
  \bibnamefont {Keitel}},\ }\bibfield  {title} {\bibinfo {title} {Extremely
  high-intensity laser interactions with fundamental quantum systems},\ }\href
  {https://doi.org/10.1103/RevModPhys.84.1177} {\bibfield  {journal} {\bibinfo
  {journal} {Rev. Mod. Phys.}\ }\textbf {\bibinfo {volume} {84}},\ \bibinfo
  {pages} {1177} (\bibinfo {year} {2012})}\BibitemShut {NoStop}%
\bibitem [{\citenamefont {Di~Piazza}\ \emph {et~al.}(2018)\citenamefont
  {Di~Piazza}, \citenamefont {Tamburini}, \citenamefont {Meuren},\ and\
  \citenamefont {Keitel}}]{DiPiazza2018}%
  \BibitemOpen
  \bibfield  {author} {\bibinfo {author} {\bibfnamefont {A.}~\bibnamefont
  {Di~Piazza}}, \bibinfo {author} {\bibfnamefont {M.}~\bibnamefont
  {Tamburini}}, \bibinfo {author} {\bibfnamefont {S.}~\bibnamefont {Meuren}},\
  and\ \bibinfo {author} {\bibfnamefont {C.~H.}\ \bibnamefont {Keitel}},\
  }\bibfield  {title} {\bibinfo {title} {Implementing nonlinear compton
  scattering beyond the local-constant-field approximation},\ }\href
  {https://doi.org/10.1103/PhysRevA.98.012134} {\bibfield  {journal} {\bibinfo
  {journal} {Phys. Rev. A}\ }\textbf {\bibinfo {volume} {98}},\ \bibinfo
  {pages} {012134} (\bibinfo {year} {2018})}\BibitemShut {NoStop}%
\bibitem [{\citenamefont {Bula}\ \emph {et~al.}(1996)\citenamefont {Bula},
  \citenamefont {McDonald}, \citenamefont {Prebys}, \citenamefont {Bamber},
  \citenamefont {Boege}, \citenamefont {Kotseroglou}, \citenamefont
  {Melissinos}, \citenamefont {Meyerhofer}, \citenamefont {Ragg}, \citenamefont
  {Burke} \emph {et~al.}}]{Bula1996}%
  \BibitemOpen
  \bibfield  {author} {\bibinfo {author} {\bibfnamefont {C.}~\bibnamefont
  {Bula}}, \bibinfo {author} {\bibfnamefont {K.~T.}\ \bibnamefont {McDonald}},
  \bibinfo {author} {\bibfnamefont {E.~J.}\ \bibnamefont {Prebys}}, \bibinfo
  {author} {\bibfnamefont {C.}~\bibnamefont {Bamber}}, \bibinfo {author}
  {\bibfnamefont {S.}~\bibnamefont {Boege}}, \bibinfo {author} {\bibfnamefont
  {T.}~\bibnamefont {Kotseroglou}}, \bibinfo {author} {\bibfnamefont {A.~C.}\
  \bibnamefont {Melissinos}}, \bibinfo {author} {\bibfnamefont {D.~D.}\
  \bibnamefont {Meyerhofer}}, \bibinfo {author} {\bibfnamefont
  {W.}~\bibnamefont {Ragg}}, \bibinfo {author} {\bibfnamefont {D.~L.}\
  \bibnamefont {Burke}}, \emph {et~al.},\ }\bibfield  {title} {\bibinfo {title}
  {Observation of nonlinear effects in {Compton} scattering},\ }\href
  {https://doi.org/10.1103/PhysRevLett.76.3116} {\bibfield  {journal} {\bibinfo
   {journal} {Phys. Rev. Lett.}\ }\textbf {\bibinfo {volume} {76}},\ \bibinfo
  {pages} {3116} (\bibinfo {year} {1996})}\BibitemShut {NoStop}%
\bibitem [{\citenamefont {Bamber}\ \emph {et~al.}(1999)\citenamefont {Bamber},
  \citenamefont {Boege}, \citenamefont {Koffas}, \citenamefont {Kotseroglou},
  \citenamefont {Melissinos}, \citenamefont {Meyerhofer}, \citenamefont {Reis},
  \citenamefont {Ragg}, \citenamefont {Bula}, \citenamefont {McDonald} \emph
  {et~al.}}]{Bamber1999}%
  \BibitemOpen
  \bibfield  {author} {\bibinfo {author} {\bibfnamefont {C.}~\bibnamefont
  {Bamber}}, \bibinfo {author} {\bibfnamefont {S.~J.}\ \bibnamefont {Boege}},
  \bibinfo {author} {\bibfnamefont {T.}~\bibnamefont {Koffas}}, \bibinfo
  {author} {\bibfnamefont {T.}~\bibnamefont {Kotseroglou}}, \bibinfo {author}
  {\bibfnamefont {A.~C.}\ \bibnamefont {Melissinos}}, \bibinfo {author}
  {\bibfnamefont {D.~D.}\ \bibnamefont {Meyerhofer}}, \bibinfo {author}
  {\bibfnamefont {D.~A.}\ \bibnamefont {Reis}}, \bibinfo {author}
  {\bibfnamefont {W.}~\bibnamefont {Ragg}}, \bibinfo {author} {\bibfnamefont
  {C.}~\bibnamefont {Bula}}, \bibinfo {author} {\bibfnamefont {K.~T.}\
  \bibnamefont {McDonald}}, \emph {et~al.},\ }\bibfield  {title} {\bibinfo
  {title} {Studies of nonlinear {QED} in collisions of 46.6 {GeV} electrons
  with intense laser pulses},\ }\href
  {https://doi.org/10.1103/PhysRevD.60.092004} {\bibfield  {journal} {\bibinfo
  {journal} {Phys. Rev. D}\ }\textbf {\bibinfo {volume} {60}},\ \bibinfo
  {pages} {092004} (\bibinfo {year} {1999})}\BibitemShut {NoStop}%
\bibitem [{\citenamefont {Cavanagh}\ \emph {et~al.}(2023)\citenamefont
  {Cavanagh}, \citenamefont {Fleck}, \citenamefont {Streeter}, \citenamefont
  {Gerstmayr}, \citenamefont {Dickson}, \citenamefont {Ballage}, \citenamefont
  {Cadas}, \citenamefont {Calvin}, \citenamefont {Dobosz~Dufr\'enoy},
  \citenamefont {Moulanier}, \citenamefont {Romagnani}, \citenamefont
  {Vasilovici}, \citenamefont {Whitehead}, \citenamefont {Specka},
  \citenamefont {Cros},\ and\ \citenamefont {Sarri}}]{Cavanagh2023}%
  \BibitemOpen
  \bibfield  {author} {\bibinfo {author} {\bibfnamefont {N.}~\bibnamefont
  {Cavanagh}}, \bibinfo {author} {\bibfnamefont {K.}~\bibnamefont {Fleck}},
  \bibinfo {author} {\bibfnamefont {M.~J.~V.}\ \bibnamefont {Streeter}},
  \bibinfo {author} {\bibfnamefont {E.}~\bibnamefont {Gerstmayr}}, \bibinfo
  {author} {\bibfnamefont {L.~T.}\ \bibnamefont {Dickson}}, \bibinfo {author}
  {\bibfnamefont {C.}~\bibnamefont {Ballage}}, \bibinfo {author} {\bibfnamefont
  {R.}~\bibnamefont {Cadas}}, \bibinfo {author} {\bibfnamefont
  {L.}~\bibnamefont {Calvin}}, \bibinfo {author} {\bibfnamefont
  {S.}~\bibnamefont {Dobosz~Dufr\'enoy}}, \bibinfo {author} {\bibfnamefont
  {I.}~\bibnamefont {Moulanier}}, \bibinfo {author} {\bibfnamefont
  {L.}~\bibnamefont {Romagnani}}, \bibinfo {author} {\bibfnamefont
  {O.}~\bibnamefont {Vasilovici}}, \bibinfo {author} {\bibfnamefont
  {A.}~\bibnamefont {Whitehead}}, \bibinfo {author} {\bibfnamefont
  {A.}~\bibnamefont {Specka}}, \bibinfo {author} {\bibfnamefont
  {B.}~\bibnamefont {Cros}},\ and\ \bibinfo {author} {\bibfnamefont
  {G.}~\bibnamefont {Sarri}},\ }\bibfield  {title} {\bibinfo {title}
  {Experimental characterization of a single-shot spectrometer for high-flux,
  {GeV}-scale gamma-ray beams},\ }\href
  {https://doi.org/10.1103/PhysRevResearch.5.043046} {\bibfield  {journal}
  {\bibinfo  {journal} {Phys. Rev. Res.}\ }\textbf {\bibinfo {volume} {5}},\
  \bibinfo {pages} {043046} (\bibinfo {year} {2023})}\BibitemShut {NoStop}%
\bibitem [{\citenamefont {Cole}\ \emph {et~al.}(2018)\citenamefont {Cole} \emph
  {et~al.}}]{Cole2018}%
  \BibitemOpen
  \bibfield  {author} {\bibinfo {author} {\bibfnamefont {J.~M.}\ \bibnamefont
  {Cole}} \emph {et~al.},\ }\bibfield  {title} {\bibinfo {title} {Experimental
  evidence of radiation reaction in the collision of a high-intensity laser
  pulse with a laser-wakefield accelerated electron beam},\ }\href
  {https://doi.org/10.1103/PhysRevX.8.011020} {\bibfield  {journal} {\bibinfo
  {journal} {Phys. Rev. X}\ }\textbf {\bibinfo {volume} {8}},\ \bibinfo {pages}
  {011020} (\bibinfo {year} {2018})}\BibitemShut {NoStop}%
\bibitem [{\citenamefont {Poder}\ \emph {et~al.}(2018)\citenamefont {Poder}
  \emph {et~al.}}]{Poder2018}%
  \BibitemOpen
  \bibfield  {author} {\bibinfo {author} {\bibfnamefont {K.}~\bibnamefont
  {Poder}} \emph {et~al.},\ }\bibfield  {title} {\bibinfo {title} {Experimental
  signatures of the quantum nature of radiation reaction in the field of an
  ultraintense laser},\ }\href {https://doi.org/10.1103/PhysRevX.8.031004}
  {\bibfield  {journal} {\bibinfo  {journal} {Phys. Rev. X}\ }\textbf {\bibinfo
  {volume} {8}},\ \bibinfo {pages} {031004} (\bibinfo {year}
  {2018})}\BibitemShut {NoStop}%
\bibitem [{\citenamefont {Berson}(1969)}]{Berson1969}%
  \BibitemOpen
  \bibfield  {author} {\bibinfo {author} {\bibfnamefont {I.}~\bibnamefont
  {Berson}},\ }\bibfield  {title} {\bibinfo {title} {Electron in the quantized
  field of a monochromatic electromagnetic wave},\ }\href@noop {} {\bibfield
  {journal} {\bibinfo  {journal} {Sov. Phys. JETP}\ }\textbf {\bibinfo {volume}
  {29}},\ \bibinfo {pages} {871} (\bibinfo {year} {1969})}\BibitemShut
  {NoStop}%
\bibitem [{\citenamefont {Bergou}\ and\ \citenamefont
  {Varr{\'o}}(1981{\natexlab{a}})}]{BergouVarro1981Nonrel}%
  \BibitemOpen
  \bibfield  {author} {\bibinfo {author} {\bibfnamefont {J.}~\bibnamefont
  {Bergou}}\ and\ \bibinfo {author} {\bibfnamefont {S.}~\bibnamefont
  {Varr{\'o}}},\ }\bibfield  {title} {\bibinfo {title} {Nonlinear scattering
  processes in the presence of a quantised radiation field. i. non-relativistic
  treatment},\ }\href@noop {} {\bibfield  {journal} {\bibinfo  {journal} {J.
  Phys. A: Math. Gen.}\ }\textbf {\bibinfo {volume} {14}},\ \bibinfo {pages}
  {1469} (\bibinfo {year} {1981}{\natexlab{a}})}\BibitemShut {NoStop}%
\bibitem [{\citenamefont {Bergou}\ and\ \citenamefont
  {Varr{\'o}}(1981{\natexlab{b}})}]{BergouVarro1981Rel}%
  \BibitemOpen
  \bibfield  {author} {\bibinfo {author} {\bibfnamefont {J.}~\bibnamefont
  {Bergou}}\ and\ \bibinfo {author} {\bibfnamefont {S.}~\bibnamefont
  {Varr{\'o}}},\ }\bibfield  {title} {\bibinfo {title} {Nonlinear scattering
  processes in the presence of a quantised radiation field. ii. relativistic
  treatment},\ }\href@noop {} {\bibfield  {journal} {\bibinfo  {journal} {J.
  Phys. A: Math. Gen.}\ }\textbf {\bibinfo {volume} {14}},\ \bibinfo {pages}
  {2281} (\bibinfo {year} {1981}{\natexlab{b}})}\BibitemShut {NoStop}%
\bibitem [{\citenamefont {Guo}\ and\ \citenamefont
  {{\AA}berg}(1988)}]{GuoAberg1988}%
  \BibitemOpen
  \bibfield  {author} {\bibinfo {author} {\bibfnamefont {D.-S.}\ \bibnamefont
  {Guo}}\ and\ \bibinfo {author} {\bibfnamefont {T.}~\bibnamefont
  {{\AA}berg}},\ }\bibfield  {title} {\bibinfo {title} {Quantum
  electrodynamical approach to multiphoton ionisation in the high-intensity
  field},\ }\href@noop {} {\bibfield  {journal} {\bibinfo  {journal} {J. Phys.
  A: Math. Gen.}\ }\textbf {\bibinfo {volume} {21}},\ \bibinfo {pages} {4577}
  (\bibinfo {year} {1988})}\BibitemShut {NoStop}%
\bibitem [{\citenamefont {Gonoskov}\ \emph {et~al.}(2016)\citenamefont
  {Gonoskov}, \citenamefont {Tsatrafyllis}, \citenamefont {Kominis},\ and\
  \citenamefont {Tzallas}}]{Gonoskov2016}%
  \BibitemOpen
  \bibfield  {author} {\bibinfo {author} {\bibfnamefont {I.~A.}\ \bibnamefont
  {Gonoskov}}, \bibinfo {author} {\bibfnamefont {N.}~\bibnamefont
  {Tsatrafyllis}}, \bibinfo {author} {\bibfnamefont {I.~K.}\ \bibnamefont
  {Kominis}},\ and\ \bibinfo {author} {\bibfnamefont {P.}~\bibnamefont
  {Tzallas}},\ }\bibfield  {title} {\bibinfo {title} {Quantum optical
  signatures in strong-field laser physics: Infrared photon counting in
  high-order-harmonic generation},\ }\href {https://doi.org/10.1038/srep32821}
  {\bibfield  {journal} {\bibinfo  {journal} {Sci. Rep.}\ }\textbf {\bibinfo
  {volume} {6}},\ \bibinfo {pages} {32821} (\bibinfo {year}
  {2016})}\BibitemShut {NoStop}%
\bibitem [{\citenamefont {Gombk{\"o}t{\H{o}}}\ \emph
  {et~al.}(2020)\citenamefont {Gombk{\"o}t{\H{o}}}, \citenamefont {Varr{\'o}},
  \citenamefont {Mati},\ and\ \citenamefont {F{\"o}ldi}}]{Gombkoto2020}%
  \BibitemOpen
  \bibfield  {author} {\bibinfo {author} {\bibfnamefont {{\'A}.}~\bibnamefont
  {Gombk{\"o}t{\H{o}}}}, \bibinfo {author} {\bibfnamefont {S.}~\bibnamefont
  {Varr{\'o}}}, \bibinfo {author} {\bibfnamefont {P.}~\bibnamefont {Mati}},\
  and\ \bibinfo {author} {\bibfnamefont {P.}~\bibnamefont {F{\"o}ldi}},\
  }\bibfield  {title} {\bibinfo {title} {High-order harmonic generation as
  induced by a quantized field: Phase-space picture},\ }\href
  {https://doi.org/10.1103/PhysRevA.101.013418} {\bibfield  {journal} {\bibinfo
   {journal} {Phys. Rev. A}\ }\textbf {\bibinfo {volume} {101}},\ \bibinfo
  {pages} {013418} (\bibinfo {year} {2020})}\BibitemShut {NoStop}%
\bibitem [{\citenamefont {Gorlach}\ \emph {et~al.}(2020)\citenamefont
  {Gorlach}, \citenamefont {Neufeld}, \citenamefont {Rivera}, \citenamefont
  {Cohen},\ and\ \citenamefont {Kaminer}}]{Gorlach2020}%
  \BibitemOpen
  \bibfield  {author} {\bibinfo {author} {\bibfnamefont {A.}~\bibnamefont
  {Gorlach}}, \bibinfo {author} {\bibfnamefont {O.}~\bibnamefont {Neufeld}},
  \bibinfo {author} {\bibfnamefont {N.}~\bibnamefont {Rivera}}, \bibinfo
  {author} {\bibfnamefont {O.}~\bibnamefont {Cohen}},\ and\ \bibinfo {author}
  {\bibfnamefont {I.}~\bibnamefont {Kaminer}},\ }\bibfield  {title} {\bibinfo
  {title} {The quantum-optical nature of high harmonic generation},\ }\href
  {https://doi.org/10.1038/s41467-020-18218-w} {\bibfield  {journal} {\bibinfo
  {journal} {Nat. Commun.}\ }\textbf {\bibinfo {volume} {11}},\ \bibinfo
  {pages} {4598} (\bibinfo {year} {2020})}\BibitemShut {NoStop}%
\bibitem [{\citenamefont {Varr{\'o}}(2021)}]{Varro2021}%
  \BibitemOpen
  \bibfield  {author} {\bibinfo {author} {\bibfnamefont {S.}~\bibnamefont
  {Varr{\'o}}},\ }\bibfield  {title} {\bibinfo {title} {Quantum optical aspects
  of high-harmonic generation},\ }\href
  {https://doi.org/10.3390/photonics8070269} {\bibfield  {journal} {\bibinfo
  {journal} {Photonics}\ }\textbf {\bibinfo {volume} {8}},\ \bibinfo {pages}
  {269} (\bibinfo {year} {2021})}\BibitemShut {NoStop}%
\bibitem [{\citenamefont {Varr{\'o}}(2022)}]{Varro2022}%
  \BibitemOpen
  \bibfield  {author} {\bibinfo {author} {\bibfnamefont {S.}~\bibnamefont
  {Varr{\'o}}},\ }\bibfield  {title} {\bibinfo {title} {Coherent and incoherent
  superposition of transition matrix elements of the squeezing operator},\
  }\href {https://doi.org/10.1088/1367-2630/ac6b4d} {\bibfield  {journal}
  {\bibinfo  {journal} {New J. Phys.}\ }\textbf {\bibinfo {volume} {24}},\
  \bibinfo {pages} {053035} (\bibinfo {year} {2022})}\BibitemShut {NoStop}%
\bibitem [{\citenamefont {Gorlach}\ \emph {et~al.}(2023)\citenamefont
  {Gorlach}, \citenamefont {Even~Tzur}, \citenamefont {Birk}, \citenamefont
  {Kr{\"u}ger}, \citenamefont {Rivera}, \citenamefont {Cohen},\ and\
  \citenamefont {Kaminer}}]{Gorlach2023}%
  \BibitemOpen
  \bibfield  {author} {\bibinfo {author} {\bibfnamefont {A.}~\bibnamefont
  {Gorlach}}, \bibinfo {author} {\bibfnamefont {M.}~\bibnamefont {Even~Tzur}},
  \bibinfo {author} {\bibfnamefont {M.}~\bibnamefont {Birk}}, \bibinfo {author}
  {\bibfnamefont {M.}~\bibnamefont {Kr{\"u}ger}}, \bibinfo {author}
  {\bibfnamefont {N.}~\bibnamefont {Rivera}}, \bibinfo {author} {\bibfnamefont
  {O.}~\bibnamefont {Cohen}},\ and\ \bibinfo {author} {\bibfnamefont
  {I.}~\bibnamefont {Kaminer}},\ }\bibfield  {title} {\bibinfo {title}
  {High-harmonic generation driven by quantum light},\ }\href
  {https://doi.org/10.1038/s41567-023-02127-y} {\bibfield  {journal} {\bibinfo
  {journal} {Nat. Phys.}\ }\textbf {\bibinfo {volume} {19}},\ \bibinfo {pages}
  {1689} (\bibinfo {year} {2023})}\BibitemShut {NoStop}%
\bibitem [{\citenamefont {Even~Tzur}\ \emph {et~al.}(2024)\citenamefont
  {Even~Tzur}, \citenamefont {Birk}, \citenamefont {Gorlach}, \citenamefont
  {Kaminer}, \citenamefont {Kr{\"u}ger},\ and\ \citenamefont
  {Cohen}}]{EvenTzur2024PRR}%
  \BibitemOpen
  \bibfield  {author} {\bibinfo {author} {\bibfnamefont {M.}~\bibnamefont
  {Even~Tzur}}, \bibinfo {author} {\bibfnamefont {M.}~\bibnamefont {Birk}},
  \bibinfo {author} {\bibfnamefont {A.}~\bibnamefont {Gorlach}}, \bibinfo
  {author} {\bibfnamefont {I.}~\bibnamefont {Kaminer}}, \bibinfo {author}
  {\bibfnamefont {M.}~\bibnamefont {Kr{\"u}ger}},\ and\ \bibinfo {author}
  {\bibfnamefont {O.}~\bibnamefont {Cohen}},\ }\bibfield  {title} {\bibinfo
  {title} {Generation of squeezed high-order harmonics},\ }\href
  {https://doi.org/10.1103/PhysRevResearch.6.033079} {\bibfield  {journal}
  {\bibinfo  {journal} {Phys. Rev. Res.}\ }\textbf {\bibinfo {volume} {6}},\
  \bibinfo {pages} {033079} (\bibinfo {year} {2024})}\BibitemShut {NoStop}%
\bibitem [{\citenamefont {Rasputnyi}\ \emph {et~al.}(2024)\citenamefont
  {Rasputnyi}, \citenamefont {Chen}, \citenamefont {Birk}, \citenamefont
  {Cohen}, \citenamefont {Kaminer}, \citenamefont {Kr{\"u}ger}, \citenamefont
  {Seletskiy}, \citenamefont {Chekhova},\ and\ \citenamefont
  {Tani}}]{Rasputnyi2024}%
  \BibitemOpen
  \bibfield  {author} {\bibinfo {author} {\bibfnamefont {A.}~\bibnamefont
  {Rasputnyi}}, \bibinfo {author} {\bibfnamefont {Z.}~\bibnamefont {Chen}},
  \bibinfo {author} {\bibfnamefont {M.}~\bibnamefont {Birk}}, \bibinfo {author}
  {\bibfnamefont {O.}~\bibnamefont {Cohen}}, \bibinfo {author} {\bibfnamefont
  {I.}~\bibnamefont {Kaminer}}, \bibinfo {author} {\bibfnamefont
  {M.}~\bibnamefont {Kr{\"u}ger}}, \bibinfo {author} {\bibfnamefont
  {D.}~\bibnamefont {Seletskiy}}, \bibinfo {author} {\bibfnamefont
  {M.}~\bibnamefont {Chekhova}},\ and\ \bibinfo {author} {\bibfnamefont
  {F.}~\bibnamefont {Tani}},\ }\bibfield  {title} {\bibinfo {title}
  {High-harmonic generation by a bright squeezed vacuum},\ }\href
  {https://doi.org/10.1038/s41567-024-02659-x} {\bibfield  {journal} {\bibinfo
  {journal} {Nat. Phys.}\ }\textbf {\bibinfo {volume} {20}},\ \bibinfo {pages}
  {1960} (\bibinfo {year} {2024})}\BibitemShut {NoStop}%
\bibitem [{\citenamefont {Even~Tzur}\ and\ \citenamefont
  {Cohen}(2024)}]{EvenTzurCohen2024}%
  \BibitemOpen
  \bibfield  {author} {\bibinfo {author} {\bibfnamefont {M.}~\bibnamefont
  {Even~Tzur}}\ and\ \bibinfo {author} {\bibfnamefont {O.}~\bibnamefont
  {Cohen}},\ }\bibfield  {title} {\bibinfo {title} {Motion of charged particles
  in bright squeezed vacuum},\ }\href
  {https://doi.org/10.1038/s41377-024-01381-w} {\bibfield  {journal} {\bibinfo
  {journal} {Light Sci. Appl.}\ }\textbf {\bibinfo {volume} {13}},\ \bibinfo
  {pages} {41} (\bibinfo {year} {2024})}\BibitemShut {NoStop}%
\bibitem [{\citenamefont {Qu}\ and\ \citenamefont
  {Fisch}(2024)}]{Qu_PRE_entangle24}%
  \BibitemOpen
  \bibfield  {author} {\bibinfo {author} {\bibfnamefont {K.}~\bibnamefont
  {Qu}}\ and\ \bibinfo {author} {\bibfnamefont {N.~J.}\ \bibnamefont {Fisch}},\
  }\bibfield  {title} {\bibinfo {title} {{Producing entangled photon pairs and
  quantum squeezed states in plasmas}},\ }\href
  {https://doi.org/10.1103/PhysRevE.110.065211} {\bibfield  {journal} {\bibinfo
   {journal} {Phys. Rev. E}\ }\textbf {\bibinfo {volume} {110}},\ \bibinfo
  {pages} {065211} (\bibinfo {year} {2024})}\BibitemShut {NoStop}%
\bibitem [{\citenamefont {Qu}\ and\ \citenamefont
  {Fisch}(2025)}]{Qu_squeezing2025}%
  \BibitemOpen
  \bibfield  {author} {\bibinfo {author} {\bibfnamefont {K.}~\bibnamefont
  {Qu}}\ and\ \bibinfo {author} {\bibfnamefont {N.~J.}\ \bibnamefont {Fisch}},\
  }\bibfield  {title} {\bibinfo {title} {Ultra-strong quantum squeezing
  mediated by plasma waves},\ }\href {https://arxiv.org/abs/2507.12288} {\
  (\bibinfo {year} {2025})},\ \Eprint {https://arxiv.org/abs/2507.12288}
  {arXiv:2507.12288 [physics.plasm-ph]} \BibitemShut {NoStop}%
\bibitem [{\citenamefont {Khalaf}\ and\ \citenamefont
  {Kaminer}(2023)}]{KhalafKaminer2023}%
  \BibitemOpen
  \bibfield  {author} {\bibinfo {author} {\bibfnamefont {M.}~\bibnamefont
  {Khalaf}}\ and\ \bibinfo {author} {\bibfnamefont {I.}~\bibnamefont
  {Kaminer}},\ }\bibfield  {title} {\bibinfo {title} {{Compton} scattering
  driven by intense quantum light},\ }\href
  {https://doi.org/10.1126/sciadv.ade0932} {\bibfield  {journal} {\bibinfo
  {journal} {Sci. Adv.}\ }\textbf {\bibinfo {volume} {9}},\ \bibinfo {pages}
  {eade0932} (\bibinfo {year} {2023})}\BibitemShut {NoStop}%
\bibitem [{\citenamefont {Seipt}\ \emph {et~al.}(2017)\citenamefont {Seipt},
  \citenamefont {Heinzl}, \citenamefont {Marklund},\ and\ \citenamefont
  {Bulanov}}]{SeiptHeinzlMarklundBulanov2017}%
  \BibitemOpen
  \bibfield  {author} {\bibinfo {author} {\bibfnamefont {D.}~\bibnamefont
  {Seipt}}, \bibinfo {author} {\bibfnamefont {T.}~\bibnamefont {Heinzl}},
  \bibinfo {author} {\bibfnamefont {M.}~\bibnamefont {Marklund}},\ and\
  \bibinfo {author} {\bibfnamefont {S.~S.}\ \bibnamefont {Bulanov}},\
  }\bibfield  {title} {\bibinfo {title} {{Depletion of Intense Fields}},\
  }\href {https://doi.org/10.1103/PhysRevLett.118.154803} {\bibfield  {journal}
  {\bibinfo  {journal} {Phys. Rev. Lett.}\ }\textbf {\bibinfo {volume} {118}},\
  \bibinfo {pages} {154803} (\bibinfo {year} {2017})}\BibitemShut {NoStop}%
\bibitem [{\citenamefont {Ilderton}\ and\ \citenamefont
  {Seipt}(2018)}]{IldertonSeipt2018}%
  \BibitemOpen
  \bibfield  {author} {\bibinfo {author} {\bibfnamefont {A.}~\bibnamefont
  {Ilderton}}\ and\ \bibinfo {author} {\bibfnamefont {D.}~\bibnamefont
  {Seipt}},\ }\bibfield  {title} {\bibinfo {title} {Backreaction on background
  fields: a coherent state approach},\ }\href
  {https://doi.org/10.1103/PhysRevD.97.016007} {\bibfield  {journal} {\bibinfo
  {journal} {Phys. Rev. D}\ }\textbf {\bibinfo {volume} {97}},\ \bibinfo
  {pages} {016007} (\bibinfo {year} {2018})}\BibitemShut {NoStop}%
\bibitem [{\citenamefont {Agarwal}(2012)}]{Agarwal_2012}%
  \BibitemOpen
  \bibfield  {author} {\bibinfo {author} {\bibfnamefont {G.~S.}\ \bibnamefont
  {Agarwal}},\ }\href@noop {} {\emph {\bibinfo {title} {{Quantum Optics}}}}\
  (\bibinfo  {publisher} {Cambridge University Press},\ \bibinfo {year}
  {2012})\BibitemShut {NoStop}%
\bibitem [{\citenamefont {Di~Piazza}\ and\ \citenamefont
  {Qu}(2026{\natexlab{a}})}]{DiPiazzaQu2026PRL}%
  \BibitemOpen
  \bibfield  {author} {\bibinfo {author} {\bibfnamefont {A.}~\bibnamefont
  {Di~Piazza}}\ and\ \bibinfo {author} {\bibfnamefont {K.}~\bibnamefont {Qu}},\
  }\bibfield  {title} {\bibinfo {title} {Control of nonlinear {Compton}
  scattering in a squeezed vacuum},\ }\href {https://doi.org/10.1103/gswv-7st8}
  {\bibfield  {journal} {\bibinfo  {journal} {Phys. Rev. Lett.}\ }\textbf
  {\bibinfo {volume} {136}},\ \bibinfo {pages} {085001} (\bibinfo {year}
  {2026}{\natexlab{a}})}\BibitemShut {NoStop}%
\bibitem [{\citenamefont {Schleich}\ and\ \citenamefont
  {Wheeler}(1987)}]{Schleich1987}%
  \BibitemOpen
  \bibfield  {author} {\bibinfo {author} {\bibfnamefont {W.}~\bibnamefont
  {Schleich}}\ and\ \bibinfo {author} {\bibfnamefont {J.~A.}\ \bibnamefont
  {Wheeler}},\ }\bibfield  {title} {\bibinfo {title} {Oscillations in photon
  distribution of squeezed states},\ }\href
  {https://doi.org/10.1364/JOSAB.4.001715} {\bibfield  {journal} {\bibinfo
  {journal} {J. Opt. Soc. Am. B}\ }\textbf {\bibinfo {volume} {4}},\ \bibinfo
  {pages} {1715} (\bibinfo {year} {1987})}\BibitemShut {NoStop}%
\bibitem [{\citenamefont {Di~Piazza}\ and\ \citenamefont
  {Qu}(2026{\natexlab{b}})}]{DiPiazzaQu2026FM}%
  \BibitemOpen
  \bibfield  {author} {\bibinfo {author} {\bibfnamefont {A.}~\bibnamefont
  {Di~Piazza}}\ and\ \bibinfo {author} {\bibfnamefont {K.}~\bibnamefont {Qu}},\
  }\bibfield  {title} {\bibinfo {title} {Nonlinear {Compton} scattering in a
  frequency-modulated field},\ }\href@noop {} {\bibfield  {journal} {\bibinfo
  {journal} {arXiv:2605.04011}\ } (\bibinfo {year} {2026}{\natexlab{b}})},\
  \Eprint {https://arxiv.org/abs/2605.04011} {arXiv:2605.04011 [quant-ph]}
  \BibitemShut {NoStop}%
\bibitem [{\citenamefont {Boca}\ and\ \citenamefont
  {Florescu}(2009)}]{BocaFlorescu2009}%
  \BibitemOpen
  \bibfield  {author} {\bibinfo {author} {\bibfnamefont {M.}~\bibnamefont
  {Boca}}\ and\ \bibinfo {author} {\bibfnamefont {V.}~\bibnamefont
  {Florescu}},\ }\bibfield  {title} {\bibinfo {title} {{Nonlinear Compton
  scattering with a laser pulse}},\ }\href
  {https://doi.org/10.1103/PhysRevA.80.053403} {\bibfield  {journal} {\bibinfo
  {journal} {Phys. Rev. A}\ }\textbf {\bibinfo {volume} {80}},\ \bibinfo
  {pages} {053403} (\bibinfo {year} {2009})}\BibitemShut {NoStop}%
\bibitem [{\citenamefont {Seipt}\ and\ \citenamefont
  {K{\"a}mpfer}(2011)}]{SeiptKampfer2011}%
  \BibitemOpen
  \bibfield  {author} {\bibinfo {author} {\bibfnamefont {D.}~\bibnamefont
  {Seipt}}\ and\ \bibinfo {author} {\bibfnamefont {B.}~\bibnamefont
  {K{\"a}mpfer}},\ }\bibfield  {title} {\bibinfo {title} {{Nonlinear Compton
  scattering of ultrashort intense laser pulses}},\ }\href
  {https://doi.org/10.1103/PhysRevA.83.022101} {\bibfield  {journal} {\bibinfo
  {journal} {Phys. Rev. A}\ }\textbf {\bibinfo {volume} {83}},\ \bibinfo
  {pages} {022101} (\bibinfo {year} {2011})}\BibitemShut {NoStop}%
\bibitem [{\citenamefont {Mackenroth}\ and\ \citenamefont
  {Di~Piazza}(2011)}]{MackenrothDiPiazza2011}%
  \BibitemOpen
  \bibfield  {author} {\bibinfo {author} {\bibfnamefont {F.}~\bibnamefont
  {Mackenroth}}\ and\ \bibinfo {author} {\bibfnamefont {A.}~\bibnamefont
  {Di~Piazza}},\ }\bibfield  {title} {\bibinfo {title} {{Nonlinear Compton
  scattering in ultrashort laser pulses}},\ }\href
  {https://doi.org/10.1103/PhysRevA.83.032106} {\bibfield  {journal} {\bibinfo
  {journal} {Phys. Rev. A}\ }\textbf {\bibinfo {volume} {83}},\ \bibinfo
  {pages} {032106} (\bibinfo {year} {2011})}\BibitemShut {NoStop}%
\end{thebibliography}%

\end{document}